\documentclass[AMA,STIX2COL]{WileyNJD-v2}

\usepackage{amsmath,amsthm,hyperref}
\usepackage{color, soul} %for highlighting
\usepackage{graphicx, caption}
\usepackage{mathtools}
\usepackage{comment}
\usepackage{xcolor}%for colours in listings
\usepackage{listings} %for writing R code in Latex
\usepackage{threeparttable} % Add this package for table notes
\usepackage{tabularx}
\usepackage{multirow}
\articletype{Research Article}%

\received{Date Month Year}
\revised{Date Month Year}
\accepted{Date Month Year}

\begin{document}

\title{Mapping between measurement scales in meta-analysis, with application to measures of body mass index in children} 

\author[1]{Annabel L Davies}
\author[1]{A E Ades}
\author[1]{Julian PT Higgins}

\authormark{A. DAVIES \textsc{et al}}

\titlemark{Mapping between measurement scales in meta-analysis, with application to measures of body mass index in children}

\address[1]{\orgdiv{Bristol Medical School}, \orgname{University of Bristol}, \orgaddress{\state{Bristol}, \country{United Kingdom}}}

\corres{Corresponding author is Annabel L Davies.\email{annabel.davies@bristol.ac.uk}}

\presentaddress{Canynge Hall, Clifton, Bristol, BS8 2PN}

%\fundingInfo{Text}
%\JELinfo{ejlje}

\abstract[Abstract]{Quantitative evidence synthesis methods aim to combine data from multiple medical trials to infer relative effects of different interventions. A challenge arises when trials report continuous outcomes on different measurement scales. To include all evidence in one coherent analysis, we require methods to `map' the outcomes onto a single scale. This is particularly challenging when trials report aggregate rather than individual data. We are motivated by a meta-analysis of interventions to prevent obesity in children. Trials report aggregate measurements of body mass index (BMI) either expressed as raw values or standardised for age and sex. We develop three methods for mapping between aggregate BMI data using known relationships between individual measurements on different scales. The first is an analytical method based on the mathematical definitions of z-scores and percentiles. The other two approaches involve sampling individual participant data on which to perform the conversions. One method is a straightforward sampling routine, while the other involves optimization with respect to the reported outcomes. In contrast to the analytical approach, these methods also have wider applicability for mapping between any pair of measurement scales with known or estimable individual-level relationships. We verify and contrast our methods using trials from our data set which report outcomes on multiple scales. We find that all methods recreate mean values with reasonable accuracy, but for standard deviations, optimization outperforms the other methods. However, the optimization method is more likely to underestimate standard deviations and is vulnerable to non-convergence.}


\jnlcitation{\cname{%
\author{A. L. Davies}, 
\author{A. E. Ades}, and
\author{J. P. T. Higgins}} (\cyear{2024}),
\ctitle{Mapping between aggregate level BMI data on different measurement scales in a meta-analysis of childhood obesity prevention interventions}, \cvol{2024;00:1--21}.}

\maketitle

\renewcommand\thefootnote{}
\footnotetext{\textbf{Abbreviations:} BMI, body mass index; zBMI, body mass index z-score; SD, standard deviation; RMSE, root mean squared error; MAE, mean absolute error.}

\renewcommand\thefootnote{\fnsymbol{footnote}}
\setcounter{footnote}{1}

\section{Introduction}

Evidence synthesis techniques, such as pairwise and network meta-analysis, combine data from multiple medical trials in order to compare different treatment options for the same condition.\cite{CochraneBook, DIAS:2018} The results of these analyses are extremely influential, and are commonly used to inform national clinical guidance regarding treatment recommendations.\cite{NICEguide} In the simplest scenario, each trial provides an estimate of the effect of a specific treatment on the same clinical outcome and the meta-analysis averages these values across the trials. A challenge facing meta-analysis occurs when trials report continuous outcomes on different measurement scales.\cite{GMD2} That is, all trials measure the same underlying construct but use different tools or test instruments. For example, if the outcome of interest is temperature, trialists may choose to measure this in Celsius, Fahrenheit or even Kelvin. In order to combine evidence from all eligible trials, we require methods to ‘map’ the outcomes onto a common scale.\cite{Lu:2014} Unlike measurements of temperature, the relationships between different clinical scales may be complex or even unknown. Mapping between these scales is particularly challenging in the common scenario where trial data is only available at the aggregate level (e.g. as means and standard deviations) rather than per participant.

This paper is motivated by a systematic review of interventions to prevent childhood obesity.\cite{Cochrane5to11, Cochrane12to18} Body mass index (BMI), which measures a person's weight relative to their height, is one of the most widely used tools for assessing obesity.\cite{Adab:2018} However, in children, BMI varies according to the child's age and sex. Therefore, obesity in children is more appropriately assessed using age- and sex- standardized BMI.\cite{deOnis:2010} A common standardization approach is to transform BMI to a z-score (zBMI) based on the distribution of BMI in an appropriate reference population.\cite{Must:2006} Using normality assumptions, the z-score can also be expressed as a percentile.

In our motivating data set, trials report aggregate-level outcomes in terms of either unstandardized BMI, zBMI or percentile BMI. Transformations between these scales are well-defined for individual participant data, but cannot be used for mapping between aggregate measurements. For example, the conversion between BMI and zBMI relies on information about an individual's age and sex, as well as about the wider population, while percentile and zBMI are related via a cumulative distribution function. Due to the lack of appropriate aggregate mapping methods, previous analyses of this, and similar data have been restricted to separate syntheses on each scale.\cite{Cochrane5to11, Cochrane12to18, Brown:2019, Hodder:2022} In this scenario only a subset of the total data informs each analysis, limiting the statistical power and precision of the results. A more desirable approach is to include all the trials in a single analysis by expressing the outcomes on a common scale. We choose zBMI as our primary outcome of interest, and therefore aim to map aggregate measurements of BMI and percentile onto this scale.

In meta-analysis, a common approach to synthesising outcomes on different scales is to measure treatment effects as `standardized mean differences', where each difference in means is divided by some scale-specific standard deviation (SD). However, this method relies on the assumption that the different scales are linear transformations of each other,\cite{Cummings:2011} which is not true for the BMI outcomes. Another recommended method is the `ratio of means' approach.\cite{Friedrich:2008} Here, effect estimates are calculated by taking the ratio between the mean outcomes in each arm, thus transforming them onto a unitless scale. This assumes a multiplicative rather than an additive model and requires that measurements on all scales take values greater than zero.\cite{GMD2} Since zBMI is defined to take positive and negative values, this approach is also not suitable for our data. More recently, Lu et al (2014)\cite{Lu:2014} developed a more sophisticated approach that simultaneously synthesizes treatment effects from multiple outcomes while estimating mapping coefficients between the scales. However, this method assumes individual-level transformations are unknown and makes strong assumptions about the form of aggregate relationships between the scales. Based on our knowledge of the different BMI scales, we seek a method that, rather than enforcing simplifying assumptions, exploits fully the known individual relationships.

In this paper we develop three methods for mapping between aggregate-level BMI data. The first two approaches involve making distributional assumptions for one of the outcome scales and sampling individual participant data on which to perform the conversions. One method performs the mapping based on sampling alone, while the other makes use of a simple optimization routine. Although we develop these methods in the context of BMI outcome scales, they also have wider applicability to mapping between any aggregate outcomes where individual-level relationships are available. In addition to these methods, we also develop an analytic approach to mapping between zBMI and percentile based on their mathematical definitions. 

We begin by describing our motivating data set in Section \ref{sec:data}. In Section \ref{sec:indiv} we define the three BMI measurement scales and describe how to convert between them for individual-level data. Next, we introduce our three aggregate mapping methods in Section \ref{sec:methods}. We begin with the analytical approach for mapping between percentile and zBMI. Then we define the more general sampling and optimization methods, which can be used to map from either BMI or percentile. We provide details of the specific algorithm in each case. In Section \ref{sec:application} we compare the accuracy of the different methods using observations from trials in our data set that report measurements on more than one scale. Finally, we summarize and discuss our results in Section \ref{sec:discuss}.

\section{Motivating data set}\label{sec:data}

Our work is motivated by a systematic review of nearly 250 randomized controlled trials of interventions to prevent obesity in children aged 5 to 18 years.\cite{Cochrane5to11, Cochrane12to18} We will refer to this as the obesity data set. 204 trials provide usable data on BMI, zBMI or percentile BMI. The outcome of interest on each scale is the mean difference (MD) between the intervention arm and reference arm in change from baseline. This is known as contrast-level data because it represents the comparison between two arms of a trial. For brevity, we will refer these data points as `contrasts'. Arm-level data, on the other hand, refer to the means and standard deviations of measurements (BMI, zBMI or percentile) in each arm at each time point. We will perform our mapping procedure on the arm-level data. The results can then be used to reconstruct the relevant contrasts for analysis.

\begin{figure*}
	%\centering
	\centerline{\includegraphics[width=1\textwidth]{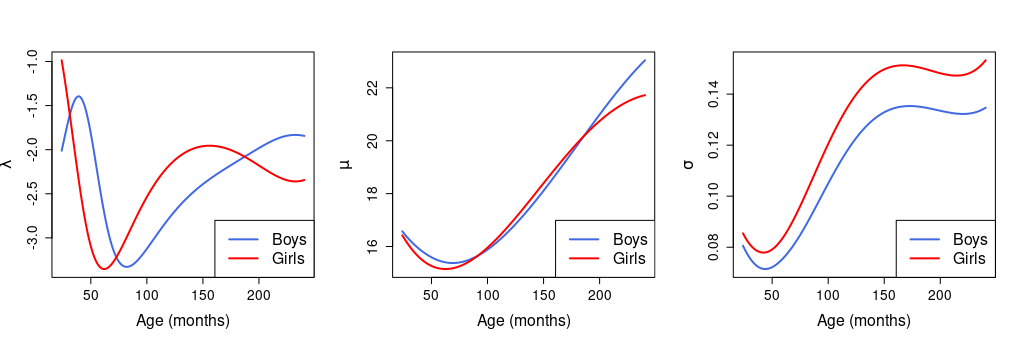}}
	\caption{CDC (Centers for Disease Control and Prevention) reference parameters $\{\lambda, \mu, \sigma\}$ as a function of age. Parameter values obtained from Reference\cite{CDC}.}
	\label{fig:CDC-LMS}
\end{figure*}

The obesity data set includes multi-arm trials and observations at multiple follow-up times, such that each trial can be associated with more than one data point. Additionally, some trials report the same contrast on more than one outcome scale. Overall, there are 182 contrasts measured on the BMI scale, 171 on the zBMI scale and 48 on the percentile scale. Due to some trials reporting multiple scales, a fully mapped data set would include 286 observed contrasts. An appropriate mapping procedure would therefore yield a much larger sample compared with analysing any one of the outcomes alone.

To test the methods we develop, we will use trials that report arm-level data on more than one outcome scale. From these trials we have 171 arm-level means measured on both the BMI and zBMI scale, 20 means measured as percentile and zBMI, and 34 values measured on all three scales. We will use these data to investigate how well the mapping procedures recreate the observed zBMI means from the reported BMI and percentile values. In these tests we will treat each data point independently, though we note that this is not strictly true since an individual trial is associated with multiple arm-level means.

\section{BMI outcomes: individual-level definitions}\label{sec:indiv}
We define $B$, $Z$, and $P$ as BMI, zBMI and percentile respectively. For each arm and time point of a particular trial, we write $\mathcal{D}_{X} = (\bar{X}, \text{SD}(X))$ for the aggregate-level data on scale $X\in \{B,Z,P\}$, where $\bar{X}$ is the reported mean and $\text{SD}(X)$ is the standard deviation on that scale. In the following, we define the relationships between these scales for a particular individual.

\subsection{BMI and zBMI}\label{sec:indiv-BMI}

The body mass index (BMI) of an individual is defined as 
\begin{align}
	B = \frac{M}{H^2},
\end{align}
where $M$ is the person's mass in kilograms (kg) and $H$ is their height in metres (m). In adults, obesity is defined when an individual's BMI crosses a certain threshold. For example, the World Health Organization (WHO) define obesity to be a BMI greater than or equal to $30 \text{ kg}/\text{m}^2$.\cite{WHOreport} In children, however, BMI varies with age and sex making it more difficult to define meaningful thresholds. Instead, weight classifications in children are defined by comparing a child's BMI with the distribution of BMI in other children of the same age and sex. This comparison is done by expressing BMI as a z-score.\cite{Must:2006}

For a normally distributed random variable, the z-score describes how many standard deviations an individual measurement is away from the mean of the distribution. It is calculated by subtracting the mean from the individual measurement and dividing by the standard deviation. This defines the z-score as a standard normal variable. However, measurements of BMI in the general population exhibit a skewed distribution meaning z-scores defined in this way do not follow the standard normal distribution.\cite{Silverman:2022} To deal with this problem, zBMI is usually calculated using the so-called `LMS method', an approach developed by Cole et al.\cite{Cole:1988, Cole:1992} as a way of defining z-scores for skewed variables using age-related reference standards.

The LMS method characterises the distribution of the variable of interest in terms of three parameters, $\{\lambda, \mu, \sigma\}$, where $\mu$ is the median of the variable's distribution, $\sigma$ is the coefficient of variation, and $\lambda$ is the Box-Cox power. The method assumes that raising the variable to the power $\lambda$ (or taking the logarithm of the variable if $\lambda=0$) yields a normal distribution. For a BMI, $B$, the LMS method defines the Z-score, $Z$, as a standard normal variable calculated via
\begin{align}\label{eq:Zbmi}
	Z = \begin{cases}
		\frac{1}{\lambda \sigma}((\frac{B}{\mu})^{\lambda}-1) & \lambda \neq 0\\
		\frac{1}{\sigma}\log \frac{B}{\mu} & \lambda = 0.
	\end{cases}
\end{align}

The distribution of BMI in children depends on age, sex, and the population of interest.\cite{deOnis:2010} Therefore, to calculate zBMI for an individual, we require the age- and sex-specific values of  $\{\lambda, \mu, \sigma\}$ for an appropriate reference population. These values have been calculated from large-scale surveys of children's heights and weights in various populations\cite{Cole:2000, Kuczmarski:2000, Kuczmarski:2002, WHO:survey:2006, deOnis:2006} and are available from national and international public health organization websites. For example, Figure \ref{fig:CDC-LMS} shows the LMS parameters for boys and girls as a function of age for the population of the United States, available through the Centers for Disease Control and Prevention (CDC).\cite{CDC} Here, we see that $\lambda<0$ for all children between the ages of 2 and 18 years. Reference charts based on surveys of international populations are available via the World Health Organization (WHO)\cite{WHO} and the International Obesity Task Force (IOTF).\cite{IOTF} As shown in Section A of the Supplementary Material, both these international references also indicate $\lambda<0$ across all ages. Therefore, Equation (\ref{eq:Zbmi}) reduces to the first line. The procedure for calculating a child's zBMI is then as follows: (i) record the child's BMI, age, and sex, (ii) choose an appropriate LMS reference chart for the relevant population, (iii) read off the values of $\{\lambda, \mu, \sigma\}$ for the age and sex of the child, (iv) substitute the parameters and the measured BMI into the first line of Equation (\ref{eq:Zbmi}).

\subsection{zBMI and percentile}

A percentile describes the probability that a random variable is less than a particular value. It is related to the z-score via the standard cumulative distribution function (cdf), $\Phi(.)$,
\begin{align}\label{eq:Perc}
	P(z) = \text{Prob}(Z<z) = \Phi(z) = \frac{1}{\sqrt{2\pi}} \int_{-\infty}^{z} \text{e}^{-x^2/2} \text{d}x.
\end{align}
Therefore, while there is a one-to-one mapping between zBMI and percentile BMI, this relationship is non-linear.

\section{Mapping between aggregate-level BMI outcomes}\label{sec:methods}
If we have measurements on one of the BMI scales for individuals of known age and sex, we can use Equations (\ref{eq:Zbmi}) and (\ref{eq:Perc}) to convert the values to either of the other two scales. However, in the obesity data set we do not have access to individual participant data, only the mean and standard deviation of measurements within each arm. Due to the complex non-linear relationships between the scales, we cannot simply apply Equations (\ref{eq:Zbmi}) and (\ref{eq:Perc}) to these aggregate data. 
Instead, in this section we develop more sophisticated methods that build on the individual-level transformations. These allow us to map aggregate BMI and percentile  data onto the zBMI scale. We label these mappings  $\mathcal{D}_B \xrightarrow{}{} \mathcal{D}_Z$ and $\mathcal{D}_P\xrightarrow{}{} \mathcal{D}_Z$ respectively.

\subsection{Method 1: Analytical}
\subsubsection{Percentile to zBMI}\label{sec:cdf}

In this section we develop an analytical method to perform the aggregate-level mapping from percentile to zBMI, $\mathcal{D}_P \xrightarrow{}{} \mathcal{D}_Z$. The method uses expressions for the expectation and variance of continuous random variables along with standard integral results from Owen (1980),\cite{Owen:1980}
\begin{align}
	\int_{-\infty}^{\infty} \Phi(a+bx) \phi(x) \text{d}x &= \Phi\left( \frac{a}{\sqrt{1+b^2}} \right), \label{eq:Owen1}\\
	\int_{-\infty}^{\infty} [\Phi(a+bx)]^2 \phi(x) \text{d}x &= \Phi\left(\frac{a}{\sqrt{1+b^2}}\right) \nonumber \\ &\hspace{10pt} - 2T\left( \frac{a}{\sqrt{1+b^2}}, \frac{1}{\sqrt{1+2b^2}} \right), \label{eq:Owen2}
\end{align}
where $\phi(.)$ is the probability density function (pdf) of the standard normal distribution and $T(h,a)$ is Owen's T function.\cite{Owen:1956}

In Section \ref{sec:indiv-BMI} we showed that zBMI is defined such that it follows a standard normal distribution in the general population. We assume it also follows a normal distribution for samples from the general population such as participants in randomized trials, $Z \sim \mathcal{N}(m_z, s_z^2)$, with some sample-specific mean $m_z$ and standard deviation $s_z$.\footnote{Throughout this article, we will use $m$ and $s$ to represent the mean and standard deviation of distributions to avoid confusion with the $\mu$ and $\sigma$ parameters in the LMS method.} Given the relationship between percentile BMI and zBMI in Equation (\ref{eq:Perc}), the expected value of percentile in this sample is then
\begin{align}
	\mathbb{E}_{Z}(P) = \mathbb{E}_{Z}(\Phi(Z)) = \int_{-\infty}^{\infty} f_{Z}(x)\Phi(x) \text{d}x,
\end{align}
where $f_Z(x)$ is the pdf of $Z$. Using the expression for the pdf of a normal distribution, we find
\begin{align}\label{eq:EP1}
	\mathbb{E}_{Z}(P) = \int_{-\infty}^{\infty}\frac{1}{s_z} \phi\left( \frac{x-m_z}{s_z} \right)\Phi(x) \text{d}x.
\end{align}
To evaluate this integral, we use the substitution $x' = \frac{x-m_z}{s_z}$ so that Equation (\ref{eq:EP1}) takes the form of the standard integral in Equation (\ref{eq:Owen1}) with $a=m_z$ and $b=s_z$. This gives the solution
\begin{align}\label{eq:EP}
	\mathbb{E}_{Z}(P) =  \Phi \left( \frac{m_z}{\sqrt{1+s_z^2}} \right).
\end{align}

Next, we derive an expression for the variance of $P$ as a function of $m_z$ and $s_z$. In terms of the expectation, the variance is defined as
\begin{align}\label{eq:VarP1}
    \text{Var}(P) = \mathbb{E}_{Z}(P^2) - \mathbb{E}_{Z}(P)^2.
\end{align}
Given that we already have an expression for $\mathbb{E}_{Z}(P)$, we must now evaluate
\begin{align}\label{eq:EPsq}
	\mathbb{E}_{Z}(P^2) = \mathbb{E}_{Z}([\Phi(Z)]^2) = \int_{-\infty}^{\infty} \frac{1}{s_z} \phi\left( \frac{x-m_z}{s_z} \right)[\Phi(x)]^2 \text{d}x.
\end{align}
Using the same substitution as before, $x' = \frac{x-m_z}{s_z}$, Equation (\ref{eq:EPsq}) takes the form of the second standard integral, Equation (\ref{eq:Owen2}), with  $a=m_z$ and $b=s_z$. From this, we obtain the result
\begin{align}\label{eq:EPsq2}
	\mathbb{E}_{Z}(P^2) = \Phi\left(\frac{m_z}{\sqrt{1+s_z^2}}\right)  - 2T\left( \frac{m_z}{\sqrt{1+s_z^2}}, \frac{1}{\sqrt{1+2s_z^2}} \right).
\end{align}
Finally, substituting Equations (\ref{eq:EP}) and (\ref{eq:EPsq2}) into Equation (\ref{eq:VarP1}) gives the following expression for the variance of $P$,
\begin{multline}\label{eq:VarP}
	\text{Var}(P) = \Phi\left(\frac{m_z}{\sqrt{1+s_z^2}}\right)  - 2T\left( \frac{m_z}{\sqrt{1+s_z^2}}, \frac{1}{\sqrt{1+2s_z^2}} \right) \\- \left(\Phi\left( \frac{m_z}{\sqrt{1+s_z^2}} \right) \right)^2.
\end{multline}
To perform the mapping $\mathcal{D}_P \xrightarrow{}{} \mathcal{D}_Z$, we set the expectation of percentile in Equation (\ref{eq:EP}) equal to the reported sample mean, $\mathbb{E}_Z(P)=\bar{P}$, and the variance in Equation (\ref{eq:VarP}) equal to the squared standard deviation, $\text{Var}(P)=\text{SD}(P)^2$. Then, we obtain the mapped dataset, $\mathcal{D}_Z$, by solving the resulting simultaneous equations for $m_z$ and $s_z$. This must be done using numerical methods, meaning the method is not strictly analytical. However, we retain the name `analytical method' to distinguish it from the subsequent methods based on sampling and optimization. 

\subsection{Method 2: Sampling}

The relationship between BMI and zBMI in Equation (\ref{eq:Zbmi}) depends implicitly on the individual's age and sex via dependence on the LMS parameters. This prevents us from deriving an analytic method to perform this mapping on the aggregate level. Instead, we propose a sampling method which can be used to map from both percentile and BMI.

The general idea, when mapping from some scale $X$ to another scale $Y$, is as follows: (i) assume a sampling distribution for scale $X$, (ii) specify the parameters of the distribution using the observed aggregate data on that scale, $\mathcal{D}_X$, (iii) sample individual participant data from this distribution, (iv) use the appropriate individual-level mapping relationship to convert the sampled data on scale $X$ to scale $Y$, (v) calculate the mean and SD of the individual-level data on scale $Y$ to define $\mathcal{D}_Y$. 

In Sections \ref{sec:samp-perc} and \ref{sec:samp-bmi} we outline the specific sampling routines to perform the mappings  $\mathcal{D}_P\xrightarrow{}{} \mathcal{D}_Z$ and $\mathcal{D}_B \xrightarrow{}{} \mathcal{D}_Z$ respectively. 

\subsubsection{Algorithm 1: Percentile to zBMI via sampling}\label{sec:samp-perc}

To map from percentile to zBMI using the sampling method, we must first assume a distribution for $P$. As shown in Equation (\ref{eq:Perc}), percentile BMI is defined as the standard normal cdf of $Z$. Given zBMI is a standard normal variable in the general population, the corresponding distribution of percentile is uniform between 0 and 1. However, to describe percentile BMI in a sample population (e.g. in a particular trial), we require a more flexible distribution. We choose the beta distribution, $\text{Beta}(\alpha, \beta)$, because it has the correct $[0,1]$ support and contains the uniform distribution as a special case when $\alpha=\beta=1$. Furthermore, as discussed in Section B of the Supplementary Material, a beta distribution for $P$ produces a normal or approximately normal distribution for $Z$ under certain parameterisations. This choice therefore aligns with our previous assumptions in Section \ref{sec:cdf} about the distribution of zBMI in a sample population.

The mean and variance of a beta random variable, $X\sim\text{Beta}(\alpha, \beta)$, with parameters $\alpha$ and $\beta$ are given by,
\begin{align}
	\mathbb{E}(X) &= \frac{\alpha}{\alpha+\beta},\\
	\text{Var}(X) &= \frac{\alpha \beta}{(\alpha+\beta)^2(\alpha+\beta+1)}. 
\end{align} 
To implement the mapping $\mathcal{D}_P \xrightarrow{} \mathcal{D}_Z$ via sampling, we obtain the $\alpha$ and $\beta$ parameters of the percentile sampling distribution by equating the observed mean and variance of the aggregate data with these expectations.  
The method then proceeds as follows:
\begin{enumerate}
	\item For $i=1,\hdots, N$:
	\begin{enumerate}
		\item[(i)] Sample percentile BMI from $P_i\sim \text{Beta}(\alpha, \beta)$ where
		\begin{align}
			\alpha &= \bar{P}^2\left( \frac{1-\bar{P}}{\text{SD}(P)^2} - \frac{1}{\bar{P}} \right),\\
			\beta &=\alpha\left( \frac{1}{\bar{P}} - 1 \right).
		\end{align}
		\item[(ii)] Calculate the corresponding value of zBMI by rearranging Equation (\ref{eq:Perc}),
		\begin{align}
			Z_i = \Phi^{-1}(P_i). 
		\end{align}
	\end{enumerate}
	\item Calculate the mean and standard deviation of the $\{Z_i\}$ to define the mapped dataset $\mathcal{D}_Z=(\bar{Z}, \text{SD}(Z))$.
\end{enumerate}

\subsubsection{Algorithm 2: BMI to zBMI via sampling}\label{sec:samp-bmi}

Silverman and Lipscombe (2022) show that BMI in the adult population follows a lognormal distribution. Given that $\lambda \neq 0$ in any BMI reference chart, this is not strictly true for children. However, the lognormal distribution is a practical choice as it is easy to sample from, provides flexibility and has the property of skew. Therefore, to set-up the sampling method for $\mathcal{D}_B \xrightarrow{} \mathcal{D}_Z$, we set the observed sample mean and SD on the BMI scale, $\mathcal{D}_B=(\bar{B}, \text{SD}(B))$, equal to the mean and (square root of) variance of a lognormal variable.

To obtain zBMI from BMI for an individual sample $i$, we require the appropriate LMS parameters, $\{\lambda, \mu, \sigma\}$. For a given reference chart, these parameters are functions of the age, $a_i$, and sex, $x_i$, of the individual. We assume trials provide aggregate information about the age and sex of its participants.  Therefore, at each iteration we sample sex (male = 1, female = 0) from a Bernoulli distribution with probability equal to the proportion of males in the trial, $p_m$. For age, we investigate both normal and uniform distributions, specified in terms of the reported mean and SD of age, $\mathcal{D}_a = (\bar{a}, \text{SD}(a))$. The sampling method for BMI is then as follows:
\begin{enumerate}
	\item For $i=1, \hdots, N$:
	\begin{enumerate}
		\item[(i)] Sample BMI from a lognormal distribution, $B_i \sim \text{LN}(m_B, s_B^2)$, where 
		\begin{align}
			m_B &= \log\left( \frac{\bar{B}}{\sqrt{\bar{B}^2+\text{SD}(B)^2}} \right),\\
			s_B^2 &=\log \left( 1+ \frac{\text{SD}(B)^2}{\bar{B}^2} \right).
		\end{align}
		\item[(ii)] Sample sex from a Bernoulli distribution, $x_i \sim \text{Bern}(p_m)$. 
		\item[(iii)] Sample age from one of the following distributions:
		\begin{enumerate}
			\item[(a)] Normal: $a_i \sim \mathcal{N}(\bar{a}, \text{SD}(a))$,
			\item[(b)] Uniform: $a_i \sim \text{Unif}(\min(a), \max(a)),$ where $\min(a) = \bar{a} - 2\text{SD}(a)$ and $\max(a) = \bar{a} + 2\text{SD}(a)$.
		\end{enumerate}
		\item[(iv)] Read off the values of the LMS parameters at age $a_i$ and sex $x_i$ from an appropriate reference chart, $\lambda(a_i, x_i), \mu(a_i, x_i), \sigma(a_i, x_i)$.
		\item[(v)] Calculate the corresponding value of zBMI via Equation (\ref{eq:Zbmi}) for $\lambda\neq 0$,
		\begin{align}
			Z_i = \frac{1}{\lambda(a_i, x_i) \sigma(a_i, x_i)}\left(\left(\frac{B_i}{\mu(a_i, x_i)}\right)^{\lambda(a_i, x_i)}-1\right).
		\end{align}
	\end{enumerate}
 	\item Calculate the mean and standard deviation of the $\{Z_i\}$ to define the mapped dataset $\mathcal{D}_Z=(\bar{Z}, \text{SD}(Z))$.
\end{enumerate}

\subsection{Method 3: Optimization}\label{sec:opt}

To map from scale $X$ to scale $Y$, the sampling method assumes, and samples from, a distribution for $X$. However, in some scenarios, it may be easier to pick a distribution and sample from the $Y$ scale. Therefore, we develop a second mapping method that uses an optimization scheme to map from $X$ to $Y$ based on sampling $Y$. In particular, we use optimization to perform the mappings $\mathcal{D}_{B} \xrightarrow{} \mathcal{D}_Z$ and $\mathcal{D}_{P} \xrightarrow{} \mathcal{D}_Z$ by sampling individual participant data on the zBMI scale.

The general idea of the optimization method, for mapping from scale $X$ to scale $Y$, is as follows: (i) Assume a distribution for scale $Y$, (ii) initialise the parameters of this distribution to some sensible values, (iii) sample individual data from this distribution, (iv) use the appropriate individual-level mapping relationship to convert sampled data on scale $Y$ to scale $X$, (v) calculate the mean and SD of the sampled data on scale $X$, (vi) compare these values to the observed data, $\mathcal{D}_X$, and adjust the parameters of the $Y$ distribution accordingly, (vii) repeat steps (iii)-(vi) until the sampled means and SDs on scale $X$ are within some specified threshold of the observed values, (viii) calculate the mean and SD of the sampled $Y$ values at the last iteration and use these to define the mapped dataset $\mathcal{D}_Y$.

For the obesity data set we are mapping to the zBMI scale and therefore, we need to assume a sampling distribution for $Z$. As in Section \ref{sec:cdf} we model zBMI in a sample population as following a normal distribution. We initialise the mean and standard deviation to 0 and 1, representing the distribution of zBMI in the general population.

To implement the method, we use a na{\"i}ve optimization scheme. At iteration $t$, we increase (or decrease) a parameter of the $Y$ distribution by a fixed amount if the difference between the estimated parameter on the $X$ scale and its corresponding observed value in $\mathcal{D}_X$ exceeds some specified threshold. In particular, we adjust the mean and SD of the zBMI distribution based on the corresponding sampled mean and SD of BMI or percentile at that iteration. However, rather than resampling the zBMI values for each updated set of parameters (step (iii) above), we use the properties of the normal distribution to update the samples ``smoothly''. First, assume that at iteration $t$ we sample $N$ values of zBMI, $\{Z_i^{(t)}; i=1,\hdots,N\}$, from a normal distribution with mean $m_z^{(t)}$ and SD $s_z^{(t)}$. Then, as shown in Appendix \ref{app:update}, if the mean and SD are increased by $\delta_m$ and $\delta_s$ respectively, the $i^{th}$ sample of zBMI can be updated via,
\begin{align}\label{eq:update}
	Z_i^{(t+1)} = \left( 1 + \frac{\delta_{s}}{s_z^{(t)}} \right) (Z_i^{(t)}- m_z^{(t)}) + (m_z^{(t)} + \delta_{m}).
\end{align}
Using this updating scheme means we only need to generate one set of $N$ zBMI samples from the initialized distribution, thus speeding up the optimization. 

\subsubsection{Algorithm 3: Percentile to zBMI via optimization}

We now specify the optimization method for the mapping $\mathcal{D}_P \xrightarrow{} \mathcal{D}_Z$. First, we choose an incremental step, $\Delta_{\text{step}}>0$, and a tolerance threshold, $\Delta_{\text{tol}}>0$. The former defines the amount by which we increase or decrease the parameters at each iteration, while the latter dictates the accuracy required before successful convergence is reached. We discuss choices for these values in Section C of the Supplement. The method then proceeds as follows:
\begin{enumerate}
	\item Initialize the parameters: $t=0$, $m_z^{(t)}=0$, $s_z^{(t)}=1$.
	\item Generate the initial sample: For $i=1,\hdots,N$
	\begin{enumerate}
		\item[(i)] Sample zBMI from $Z_i^{(t)} \sim \mathcal{N}(m_z^{(t)}, s_z^{(t)2})$.
		\item[(ii)] Calculate the equivalent percentile, $P_i^{(t)} = \Phi(Z_i^{(t)})$.
	\end{enumerate}
	\item Calculate the means and standard deviations of the $\{Z_i^{(t)}\}$ and $\{P_i^{(t)}\}$ samples. For scale $X\in\{Z,P\}$ we label these values as $\mathcal{D}_X^{(t)} = (\bar{X}^{(t)}, \text{SD}(X)^{(t)})$.
	\item Define $\delta_m$ and $\delta_s$ by comparing $\mathcal{D}_P^{(t)}$ with the observed data, $\mathcal{D}_P$:
	\begin{align*}
		\delta_m &= \begin{cases}
			\phantom{-}0 & \text{if} \hspace{10pt} |\bar{P}^{(t)} - \bar{P}| \leq \Delta_{\text{tol}}\\
			\phantom{-} \Delta_{\text{step}} & \text{if} \hspace{10pt} \phantom{|}\bar{P}^{(t)} - \bar{P}\phantom{|} < \Delta_{\text{tol}}\\
			-\Delta_{\text{step}} & \text{if} \hspace{10pt} \phantom{|}\bar{P}^{(t)} - \bar{P}\phantom{|} > \Delta_{\text{tol}}
		\end{cases}\\ \delta_s &= \begin{cases}
		\phantom{-}0 & \text{if} \hspace{10pt} |\text{SD}(P)^{(t)} - \text{SD}(P)| \leq \Delta_{\text{tol}}\\
		\phantom{-} \Delta_{\text{step}} & \text{if} \hspace{10pt} \phantom{|}\text{SD}(P)^{(t)} - \text{SD}(P)\phantom{|} < \Delta_{\text{tol}}\\
		-\Delta_{\text{step}} & \text{if} \hspace{10pt} \phantom{|}\text{SD}(P)^{(t)}- \text{SD}(P)\phantom{|} > \Delta_{\text{tol}}
	\end{cases}
	\end{align*}
	\item Update the samples: For $i=1,\hdots,N$
	\begin{enumerate}
		\item[(i)] 	$Z_i^{(t+1)} = \left( 1 + \delta_{s}/s_z^{(t)} \right) (Z_i^{(t)}- m_z^{(t)}) + (m_z^{(t)} + \delta_{m})$
		\item[(ii)]  $P_i^{(t+1)} = \Phi(Z_i^{(t+1)})$
	\end{enumerate}
	\item Update the parameters of the zBMI distribution: $m_z^{(t+1)} = m_z^{(t)} + \delta_m$ and $s_z^{(t+1)} = s_z^{(t)} + \delta_s$.
	\item Set $t=t+1$. Go to Step 3.
\end{enumerate}
The method repeats steps 3 to 7 until one of the following conditions is met:
\begin{enumerate}
	\item[C1.]  $|\bar{P}^{(t)} - \bar{P}| \leq \Delta_{\text{tol}}$ and $|\text{SD}(P)^{(t)} - \text{SD}(P)| \leq \Delta_{\text{tol}}$
	\item[C2.]The number of iterations exceeds some specified maximum, $n_{\text{max}}$.
\end{enumerate}
We then define the mapped zBMI data set using the mean and SD of the $\{Z_i^{(t)}\}$ samples at the final iteration, $\mathcal{D}_Z = \mathcal{D}_Z^{(t)}$.  Alternatively, we could estimate $\mathcal{D}_Z$ using the final values of the parameters $m_z^{(t)}$ and $s_z^{(t)}$. We refer to the former as the sample estimates and the latter as the distribution estimates. In Section \ref{sec:ZP-results} we investigate the behaviour of both. For the mapping $\mathcal{D}_P \xrightarrow{} \mathcal{D}_Z$, the sample and distribution estimates should be almost equivalent. However, as we will see below, this is not necessarily the case for the BMI mapping due to the extra approximations involved.

\subsubsection{Algorithm 4: BMI to zBMI via optimization}

The optimization method for the mapping $\mathcal{D}_B \xrightarrow{} \mathcal{D}_Z$ follows the same steps as for percentile, but with the additional sampling of age and sex in order to specify the LMS parameters:
\begin{enumerate}
	\item Initialize the parameters: $t=0$, $m_z^{(t)}=0$, $s_z^{(t)}=1$.
	\item Generate the initial sample: For $i=1,\hdots,N$
	\begin{enumerate}
		\item[(i)] Sample zBMI from $Z_i^{(t)} \sim \mathcal{N}(m_z^{(t)}, s_z^{(t)2})$.
		\item[(ii)] Sample sex from a Bernoulli distribution, $x_i \sim \text{Bern}(p_m)$.
		\item[(iii)] Sample age from one of the following distributions:
		\begin{enumerate}
			\item[(a)] Normal: $a_i \sim \mathcal{N}(\bar{a}, \text{SD}(a))$
			\item[(b)] Uniform: $a_i \sim \text{Unif}(\min(a), \max(a)),$ where $\min(a) = \bar{a} - 2\text{SD}(a)$ and $\max(a) = \bar{a} + 2\text{SD}(a)$.
		\end{enumerate}
		\item[(iv)] Read off the values of the LMS parameters at age $a_i$ and sex $x_i$ from an appropriate reference chart, $\lambda(a_i, x_i), \mu(a_i, x_i), \sigma(a_i, x_i)$.
		\item[(v)] Calculate BMI by rearranging Equation (\ref{eq:Zbmi}) for $\lambda\neq 0$,
		\begin{align}\label{eq:BfromZ}
			B_i = \mu(a_i,x_i)(1+\lambda(a_i,x_i)\sigma(a_i,x_i)Z_i)^{\frac{1}{\lambda(a_i,x_i)}}.
		\end{align}
	\end{enumerate}
	\item Calculate the means and standard deviations of the $\{Z_i^{(t)}\}$ and $\{B_i^{(t)}\}$ samples, labelled $\mathcal{D}_X^{(t)}=(\bar{X}^{(t)}, \text{SD}(X)^{(t)})$ for $X \in \{Z,B\}$. 
	\item Define $\delta_m$ and $\delta_s$ by comparing $\mathcal{D}_B^{(t)}$ to the observed data, $\mathcal{D}_B$:
	\begin{align*}
		\delta_m &= \begin{cases}
			\phantom{-}0 & \text{if} \hspace{10pt} |\bar{B}^{(t)} - \bar{B}| \leq \Delta_{\text{tol}}\\
			\phantom{-} \Delta_{\text{step}} & \text{if} \hspace{10pt} \phantom{|}\bar{B}^{(t)} - \bar{B}\phantom{|} < \Delta_{\text{tol}}\\
			-\Delta_{\text{step}} & \text{if} \hspace{10pt} \phantom{|}\bar{B}^{(t)} - \bar{B}\phantom{|} > \Delta_{\text{tol}}
		\end{cases} , \\ \delta_s &= \begin{cases}
			\phantom{-}0 & \text{if} \hspace{10pt} |\text{SD}(B)^{(t)} - \text{SD}(B)| \leq \Delta_{\text{tol}}\\
			\phantom{-} \Delta_{\text{step}} & \text{if} \hspace{10pt} \phantom{|}\text{SD}(B)^{(t)} - \text{SD}(B)\phantom{|} < \Delta_{\text{tol}}\\
			-\Delta_{\text{step}} & \text{if} \hspace{10pt} \phantom{|}\text{SD}(B)^{(t)}- \text{SD}(B)\phantom{|} > \Delta_{\text{tol}}
		\end{cases}
	\end{align*}
	\item Update the samples: For $i=1,\hdots,N$
	\begin{enumerate}
		\item[(i)] 	$Z_i^{(t+1)} = \left( 1 + \delta_{s}/s_z^{(t)} \right) (Z_i^{(t)}- m_z^{(t)}) + (m_z^{(t)} + \delta_{m})$
		\item[(ii)]  $B_i = \mu(a_i,x_i)(1+\lambda(a_i,x_i)\sigma(a_i,x_i)Z_i)^{\frac{1}{\lambda(a_i,x_i)}}$
	\end{enumerate}
	\item Update the parameters of the zBMI distribution: $m_z^{(t+1)} = m_z^{(t)} + \delta_m$ and $s_z^{(t+1)} = s_z^{(t)} + \delta_s$.
	\item Set $t=t+1$. Go to Step 3.
\end{enumerate} 
As before, steps 3 to 7 are repeated until both the sample mean and SD of BMI are within $\Delta_{\text{tol}}$ of the observed data, or the maximum number of repetitions is reached. 

\paragraph*{Aside: Limitations of the LMS method when calculating BMI from zBMI} 

The calculation of BMI from zBMI in Equation (\ref{eq:BfromZ}) involves the exponent $1/\lambda(a_i,x_i)$. In the three major reference charts (CDC, WHO and IOTF), $\lambda<-1$ for nearly all values of $(a_i,x_i)$. This means that the exponent almost always represents a root. Since we can only take the root of a positive number, the base of this exponent must be non-negative, 
\begin{align}
	1+\lambda(a_i,x_i) \sigma(a_i,x_i) Z_i > 0,
\end{align}
where $\sigma(a_i,x_i)>0$ and $Z_i \in [-\infty, \infty]$. Sampled values of zBMI are then restricted to
\begin{align}
	Z_i < \frac{-1}{\lambda(a_i,x_i) \sigma(a_i,x_i)}.
\end{align}
The minimum value of $-1/\lambda \sigma$ across all the reference charts is 2.785, which occurs in the CDC chart for females at $a_i=240.5$ months. Therefore, for this chart, a sampled value of $Z_i$ greater than 2.785 will give an invalid value of BMI when $x_i=0$ (female) and $a_i=240.5$. The other reference charts have less restrictive cut-offs, but encounter the same issue for large values of $Z_i$ at particular ages. To deal with this problem, we truncate any problematic $Z_i$ samples to a value close to the boundary. In particular, if $Z_i>-1/\lambda(a_i,x_i) \sigma(a_i,x_i)$ we set that sample equal to $0.99\times(-1/\lambda(a_i,x_i) \sigma(a_i,x_i))$. This sets the base of the exponent equal to 0.01, which produces a large value of $B_i$.

Because we have to artificially restrict the distribution of $Z$ for certain samples, the mean and SD of the sampled values at any iteration, $\mathcal{D}_Z^{(t)} = (\bar{Z}^{(t)}, \text{SD}^{(t)}(Z))$, will not necessarily be the same as the distribution parameters, $(m_z^{(t)}, s_z^{(t)})$. In Section \ref{sec:results-bmi} we investigate the behaviour of both. However, the restriction is only necessary in a minority of cases so we do not expect it to have a major impact.

\section{Application}\label{sec:application}
\subsection{Methods}
To investigate the behaviour of the mapping methods, we compiled two test data sets from the obesity data. The first includes results from all trials that report arm-level means and SDs on both the BMI and zBMI scale (205 observations from 41 trials). The second includes all trials with arm-level data for both percentile and zBMI (54 observations from 10 trials). We refer to these as the $BZ$ and $PZ$ data sets respectively. In the following, we apply the mapping methods to either the BMI or percentile data in each test data set and compare the mapped values to the corresponding values reported on the zBMI scale.

\begin{figure*}
	\centering
	\includegraphics[width=1\linewidth]{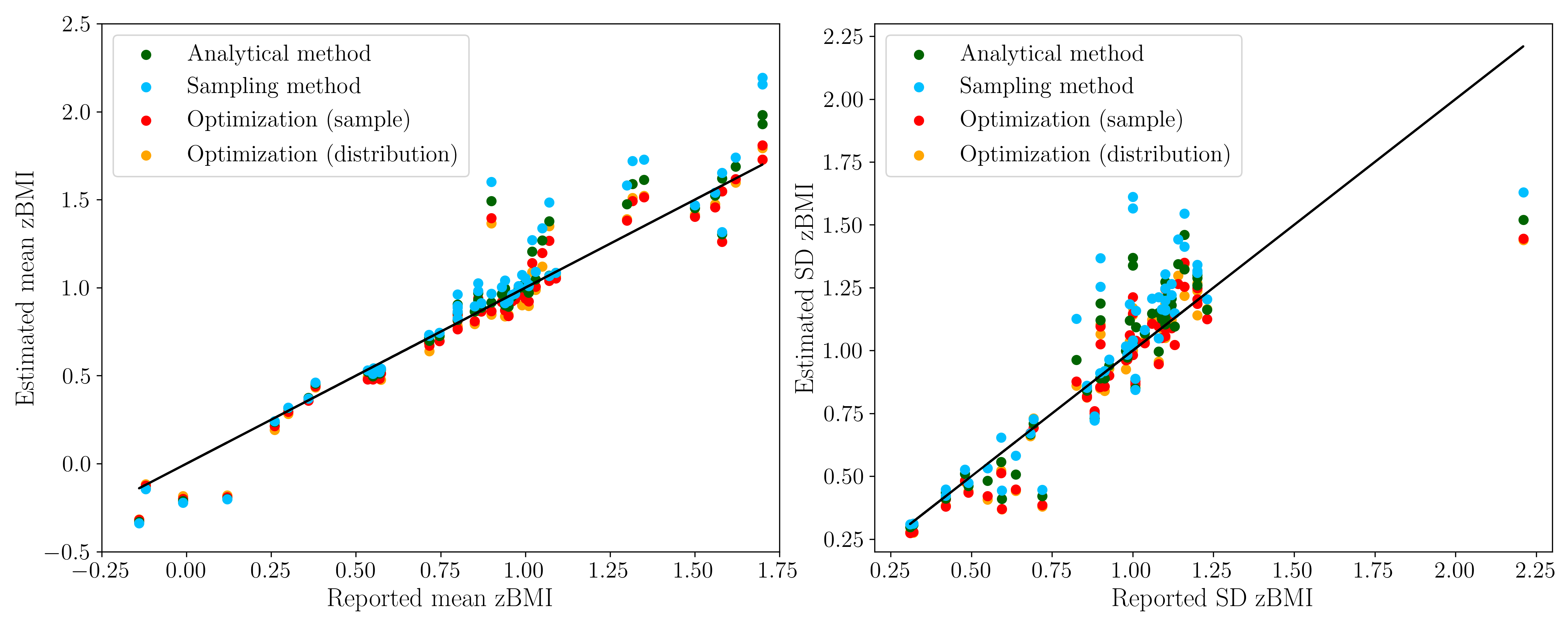}
	\caption{A plot comparing the different mapping methods for estimating aggregate zBMI data from aggregate percentile data. }
	\label{fig:ZfromP}
\end{figure*}

\subsubsection{Reconstructing arm-level data}
Most trials in our test data sets report arm-level data at baseline and follow-up for each arm of the trial. Therefore, a trial with $A_i$ arms and $T_i$ follow-up times, contributes $A_i(T_i+1)$ data points. Some trials, on the other hand, only provide arm-level data at one time point (e.g. baseline or follow-up alone) and therefore contribute $A_i$ data points. Other trials report baseline values alongside mean change from baseline (change scores). In these cases, we reconstructed follow-up means by adding the change score to the baseline score, and set the follow-up SD on each arm equal to the corresponding SD at baseline. The latter assumption is supported by the observation in other trials that SDs at different time points are approximately equal.

The obesity data include cluster-randomized trials. To construct our test data sets, we extracted raw values that had not been adjusted for clustering. When only adjusted values were available, we `unadjusted' the SDs by dividing by the square root of the design effect. Where available, we defined the design effect in terms of the intra-class correlation coefficient (ICC) reported in the trial. Otherwise, we assumed a value of $\text{ICC}=0.02$ based on ICCs reported in other trials (see our previous analyses\cite{Cochrane5to11, Cochrane12to18}).

\subsubsection{Age and sex}
To perform the mapping $\mathcal{D}_B \xrightarrow{} \mathcal{D}_Z$, we require information about the distribution of age and sex in the sample. Therefore, to construct the $BZ$ data set, we extracted the mean and SD of age (at baseline) and the proportion of male participants in each trial. For follow-up measurements, we increased the mean age by the length of follow-up. In some trials, age was reported as a range. Here, we assumed a normal distribution such that the mean is given by the mid point, and the SD is approximately a quarter of the range. When the proportion of males and females was not reported, we assumed an even split ($p_m=0.5$).

\subsubsection{Reference charts}
Mapping from BMI to zBMI also requires that we have age- and sex-specific LMS values for an appropriate reference population. Where reported, we extracted which reference chart was used by the trial authors. In all but one trial,\cite{Haerens:2006} which used LMS values specific to the Flemish population,\footnote{The Flemish reference chart cited in Haerens et al (2006)\cite{Haerens:2006} is no longer accessible. Instead we used the updated chart published by the same authors in 2009\cite{Roelants:2009}.} the trialists reported CDC, WHO or IOTF charts. Where a reference was not given, we used the CDC chart for trials set in the US and the IOTF chart for trials in any other country.

The sex-specific reference charts were downloaded as Microsoft Excel files with age tabulated against $\lambda, \mu,$ and $\sigma$. For each sample $(a_i, x_i)$, we identified the tabulated age closest to $a_i$ on the appropriate reference chart for sex $x_i$ and read off the corresponding LMS parameters. Each reference chart reports parameters for a different range of ages, at different intervals. For example, the CDC chart reports LMS parameters for ages 2 to 20 years in steps of 1 month, the IOTF chart goes from 2 to 18 years in steps of 6 months and WHO ranges from 5 to 19 years in steps of 1 month. When sampling age, we re-sampled any value of $a_i$ that was outside the range of the relevant reference chart.

\subsection{Results}

\subsubsection{zBMI from percentile}\label{sec:ZP-results}

\begin{center}
\begin{table*}
\caption{The root mean squared error (RMSE) and mean absolute error (MAE) on the estimated aggregate data (mean and SD) from the percentile mapping methods.\label{Tab:PercResults}}
\begin{tabular*}{\textwidth}{@{\extracolsep\fill}lllll@{}}
\toprule
&\multicolumn{2}{@{}l}{\textbf{Estimated mean zBMI}} & \multicolumn{2}{@{}l}{\textbf{Estimated SD of zBMI}} \\\cmidrule{2-3}\cmidrule{4-5}
\textbf{Method} & \textbf{RMSE}  & \textbf{MAE}  & {\textbf{RMSE}}  & \textbf{MAE}   \\
\midrule
Analytical & 0.147 & 0.090 & 0.303 &  0.138 \\ %\hline
	Sampling & 0.196 & 0.122 &  0.204 &  0.135  \\ %\hline
	Optimization (sample) & 0.119 & 0.079 & 0.146 & 0.089 \\ %\hline
	Optimization (distribution) & 0.119 & 0.079 & 0.148 & 0.090 \\ %\hline 
\bottomrule
\end{tabular*}
\end{table*}
\end{center}

\vspace{-25pt}
To perform the mapping $\mathcal{D}_P \xrightarrow{} \mathcal{D}_Z$, we implemented the sampling method (Algorithm 1) with $N=10^4$ samples. For the optimization method (Algorithm 3), we used $N=10^3$ samples and a maximum of $n_{\max}=5\times10^3$ iterations. To pick $\Delta_{\text{step}}$ and $\Delta_{\text{tol}}$, we explored how changes in the zBMI distribution lead to changes in the percentile distribution. We used this to propose a range of sensible values for $(\Delta_{\text{step}},\Delta_{\text{tol}})$ and investigated the corresponding convergence of the algorithm (see Section C of the Supplementary Material). Based on these preliminary tests, we chose $\Delta_{\text{step}}=0.002$ and $\Delta_{\text{tol}}=0.005$ which led to 100\% convergence.

Figure \ref{fig:ZfromP} shows estimated zBMI against reported zBMI for the different  $\mathcal{D}_P \xrightarrow{} \mathcal{D}_Z$ methods: analytical, sampling and optimization. For the latter, we show both the sample estimates, $\mathcal{D}_Z^{(t)} = (\bar{Z}^{(t)}, \text{SD}(Z)^{(t)})$, and the distribution estimates, $(m_z, s_z)$. The left panel of Figure \ref{fig:ZfromP} depicts the results for mean zBMI and the right panel shows the corresponding standard deviation.

From the Figure, it appears the different methods all exhibit similar accuracy, though sampling has slightly higher estimates on average (for both mean and SD) and optimization slightly lower. As expected, the sample and distribution estimates from the optimization method are very similar. All methods do a reasonable job of recreating mean zBMI, with most data points approximately following the `true' line. Estimated SDs are more scattered, but all take values within a realistic range.  

The right-most data point on the SD plot is underestimated by all methods and represents a notable outlier. This value corresponds to the follow-up measurement of the control arm in a trial by Brown et al. (2013).\cite{Brown:2013} On inspection we find that, on the zBMI scale, this SD measurement (2.21) is approximately twice as large as the SD reported for the other arms and time points (which range from 1.14 to 1.16). However, on the percentile scale this SD (27.3) is very similar to the other values (26.5-27.4). It is therefore not surprising that our mapping methods fail to recreate this value. This potentially points to a reporting error in the trial.

In Table \ref{Tab:PercResults} we compare the percentile mapping methods using the root mean squared error (RMSE) and the mean absolute error (MAE). These metrics quantify how accurately the mapped percentile data recreate the corresponding zBMI data reported in the trials. While all methods exhibit similar errors, the optimization method (sample estimates) produces the most accurate measurements of both mean and SD. According to these metrics, the sampling method performs worst for the mean and the analytical method performs worst for the SD.

\subsubsection{zBMI from BMI}\label{sec:results-bmi}
To map from BMI to zBMI using algorithms 2 and 4, we chose the same values of $N$ and $n_{\max}$ as we did for the percentile methods. As shown in Section C of the Supplementary Material, we explored various values of $(\Delta_{\text{step}},\Delta_{\text{tol}})$ for Algorithm 4 and they all led to comparable estimates. We show results for $\Delta_{\text{step}}=0.01$ and $\Delta_{\text{tol}}=0.1$ which gave the highest convergence rate (90\%). 

\begin{figure*}[h]
	\centering
	\includegraphics[width=1\linewidth]{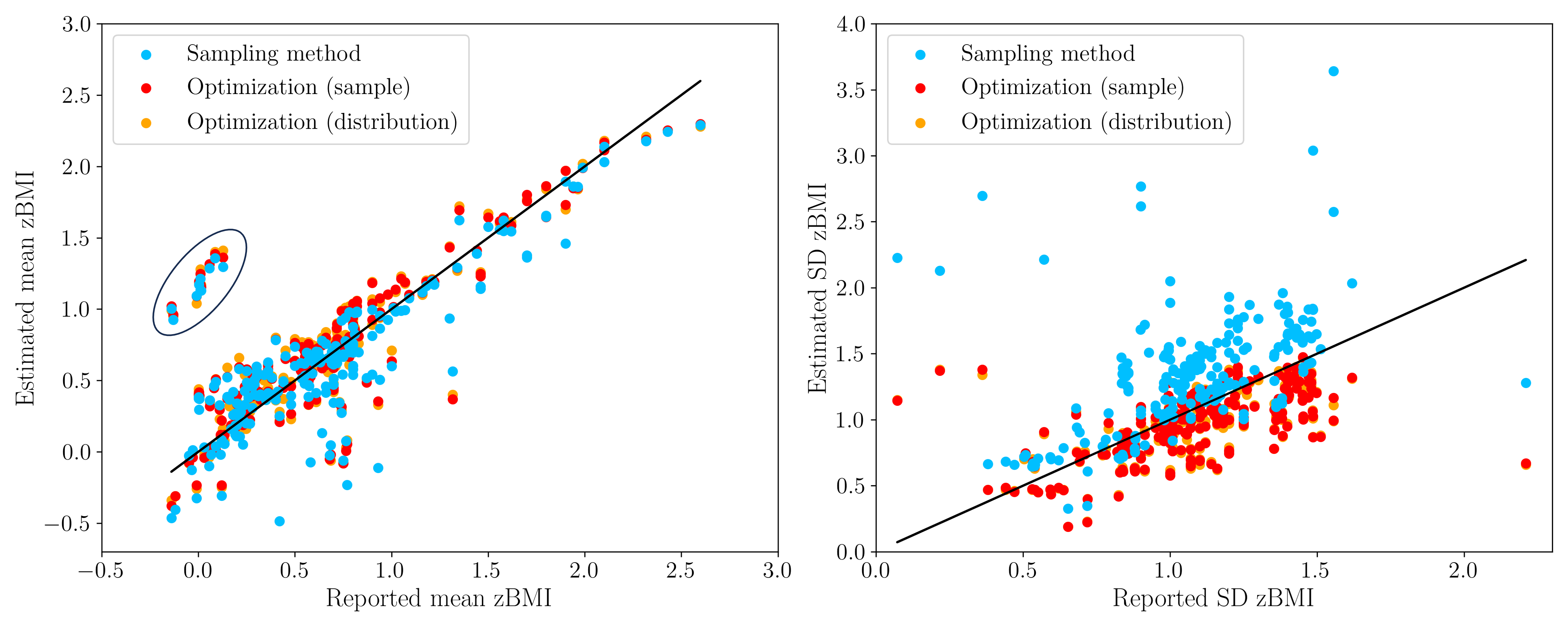}
	\caption{A plot comparing the different mapping methods for estimating aggregate zBMI data from aggregate BMI data.}
	\label{fig:ZfromB}
\end{figure*}

Figure \ref{fig:ZfromB} shows estimated zBMI against reported zBMI for the  $\mathcal{D}_B \xrightarrow{} \mathcal{D}_Z$ sampling and optimization methods. Mean values are plotted in the left panel and SDs on the right. For both methods, we present the results when age is sampled from a normal distribution. As shown in Section D of the Supplement, we observe negligible differences between the results for different age distributions.

\begin{center}
\begin{table*}[!h]%
\caption{The root mean squared error (RMSE) and mean absolute error (MAE) on the estimated aggregate data (mean and SD) from the BMI mapping methods.\label{Tab:BMIResults}}
\begin{tabular*}{\textwidth}{@{\extracolsep\fill}lllll@{}}
\toprule
&\multicolumn{2}{@{}l}{\textbf{Estimated mean zBMI}} & \multicolumn{2}{@{}l}{\textbf{Estimated SD of zBMI}} \\\cmidrule{2-3}\cmidrule{4-5}
\textbf{Method} & \textbf{RMSE}  & \textbf{MAE}  & {\textbf{RMSE}}  & \textbf{MAE}   \\
\midrule
    Sampling & 0.355 & 0.219 &  0.505 &  0.338  \\ 
	Optimization (sample) & 0.339 & 0.201 & 0.284 & 0.196 \\ 
	Optimization (distribution) & 0.341 & 0.204 & 0.283 & 0.195 \\ 
\bottomrule
\end{tabular*}
\end{table*}
\end{center}

\vspace{-37pt}
\indent On inspection of Figure \ref{fig:ZfromB}, we find that the sample and distribution estimates from the optimization method give comparable results, indicating that the restriction placed on extreme values of zBMI has a negligible effect. For mean zBMI, the sampling and optimization methods also appear to be similarly accurate. Except for a cluster of outliers (indicated in the figure), they recreate the mean values reasonably well. The circled outliers in Figure \ref{fig:ZfromB} are all from the same trial, Papadaki et al. (2010).\cite{Papadaki:2010} On inspection of this study, there is no obvious cause of the issue. The trialists state that weight status was determined using cut-offs from the IOTF, but they do not report how zBMI was calculated. It is possible that they used the simple definition of a z-score for normal distributions, rather than the LMS method. If so, this may explain the discrepancy in these results.

As shown in the right panel of Figure \ref{fig:ZfromB}, the optimization method produces more accurate estimates of SD than the sampling method, which tends to overestimate. However, the optimization method is more likely to underestimate SD. In the context of using the data in an analysis, underestimating SD is less desirable as it leads to an overweighting of the mapped values. It is possible that optimization leads to smaller SDs than the sampling method because of the truncation of extreme zBMI samples. However, the similarity of the sample and distribution estimates from the optimization method suggests that this does not have an important effect. The left-most data points on the right panel reflect reported values which seem unrealistically small for standard deviations. It is therefore not surprising they are estimated poorly by both methods. On inspection we found no issue with data extraction so this could point to a reporting error in the trial. The right-most data point is the same outlier from Brown et al. (2013) that we observed in the percentile results.

In the Supplementary Material (Section E) we explore three potential causes of inaccurate results: (i) studies that don't report a reference chart, (ii) SDs that we unadjusted for clustering, and (iii) follow-up values imputed from change scores. Excluding measurements corresponding to (i)-(iii) removes the most significant scatter on the plots for both mean and SD. To further explore the effect of reference charts, in Section F of the Supplement we plot the results separately for trials which reported each chart. For mean zBMI, no chart appears to perform systematically better or worse than any other. For standard deviation, the results associated with the CDC chart are associated with the largest scatter. 

In Section G of the Supplementary Material we show the results of the optimization method excluding any estimates which did not converge. The plot is largely similar to Figure \ref{fig:ZfromB} but, notably, does not include the right-most outlier on the SD plot. 

Finally, Table \ref{Tab:BMIResults} compares the BMI mapping methods in terms of RMSE and MAE. According to these metrics, all methods exhibit similar accuracy in the estimation of mean zBMI, but the sampling method has the largest error on SD. While these results are in agreement with the observations from Figure \ref{fig:ZfromB}, they do not reflect the tendency of the optimization method to underestimate SD.

\section{Discussion}\label{sec:discuss}

In this paper, we devised a range of methods for mapping between aggregate arm-level observations of BMI outcomes by leveraging known relationships between individual measurements on the different scales. Specifically, we proposed several methods for mapping percentile and un-standardized BMI onto zBMI. When applied to our motivating dataset, all methods yielded reasonable results, with estimated mean zBMI values aligning well with those reported in the trials. While the methods performed less effectively in reproducing standard deviations on zBMI, all estimated SDs fell within a realistic range.

Assuming a normal distribution for zBMI, we developed an analytical method based on standard integral results to map observations of mean and SD from the percentile to the zBMI scale. This deterministic approach, unlike numerical methods, remains unaffected by sample size or convergence issues, making it computationally less demanding. However, its specificity to the definition of z-scores and percentiles limits its broader applicability. Conversely, both the sampling and optimization methods offer versatility beyond our initial dataset, enabling mapping between any pair of outcomes with known individual-level relationships. To map from scale $X$ to scale $Y$, the sampling method assumes and samples from a distribution for scale $X$, while the optimization method samples from a distribution for $Y$. Therefore, the ease of sampling on each scale can guide the selection of which method to use.

All the methods we developed are sensitive to the choice of distributions for the different scales. Guided by theoretical and practical considerations, we assumed a beta distribution for percentile, a lognormal distribution for BMI, and a normal distribution for zBMI. Future work could explore the validity of these assumptions, for example using individual participant data. However, we expect that differences in trial demographics and eligibility criteria will cause variability in distributions across trials. Therefore, any choice of distribution will necessarily be an approximation. We believe the distributions we assumed are sensible choices given the information available.

The relationship between BMI and zBMI relies on information about an individual's age and sex, thus introducing additional assumptions when mapping from BMI. These include assumptions about the age distribution in the sample and the reference chart used to define the LMS parameters. Furthermore, when extracting the age-specific LMS values we used the age tabulated in the reference chart that was closest to the sampled age. A more sophisticated approach could interpolate between these values. However, we do not expect that this method had a large impact on our results. Indeed, trial authors themselves are likely to have used tabulated rather than interpolated values, meaning our approach more closely resembles the probable data generating process. Another important assumption we make is that all trial authors used the LMS method to evaluate zBMI. While this is by far the most common definition, it is possible that some authors used the standard definition of a z-score assuming a normal distribution and some measure of mean BMI. We did not find any reference to the latter, but several trials did not explain their method for calculating zBMI. Considering the number of assumptions we had to make, we believe the BMI mapping methods performed remarkably well.

Another related concern centres around potential `inaccuracies' in the LMS values themselves. Due to demographic differences and changes over time, the reported parameters may not be appropriate for a particular trial or set of trials. Our findings indicate that the BMI sampling method tends to overestimate SD, while optimization is more likely to underestimate. Since the two methods operate the LMS relation in the opposite direction ($B\xrightarrow{}Z$ for the sampling method and $Z\xrightarrow{}B$ for optimization), the observed discrepancies may stem from systematic `errors' in the LMS parameters. Indeed, we found that removing data from trials that did not report a reference chart eliminated some of the most inaccurate results. This suggests that the accuracy of the mapping methods relies on appropriate LMS parameters. Furthermore, we observed that the most extreme scatter on the SD results was associated with the CDC reference chart, pointing to a potential issue with these values. In future work, we plan to investigate these ideas further, for example examining whether variations in accuracy are related to the age or average BMI of the population. We will also explore possible methods for adjusting the LMS parameters to use in specific populations.

In our application to the motivating dataset, the RMSE and MAE metrics slightly favoured the optimization method over the sampling method. However, optimization is sensitive to convergence and may underestimate standard deviation when mapping from BMI. While the sampling method gives less accurate estimates of SD from BMI, it leans towards conservatism, reducing the risk of overweighting mapped data in subsequent analyses. To map from BMI to zBMI, the sampling method samples values of BMI while the optimization method samples zBMI. Both methods also sample age and sex, and do this independently from the sampling of the outcome scale. Given the dependence of BMI on age and sex, the sampling method might be improved by constructing an appropriate joint distribution for these variables. On the other hand, independence from age and sex is a more reasonable assumption for zBMI, which may explain the superior performance of the optimization method. 

To validate and compare our methods, we relied on aggregate data reported in trials, which are susceptible to misreporting and misinterpretation (as highlighted by some outliers in our results). Thus, small differences in RMSE and MAE should be interpreted cautiously. In future work, individual participant data might facilitate more accurate assessments of the methods.

Our optimization scheme, while largely effective, employed a na{\"i}ve approach of incrementing or decrementing parameters by a fixed amount at each iteration. Additionally, we assumed that the mean of the percentile or BMI samples was only influenced by the mean of the zBMI distribution and similarly, that the SD of the samples depended only on the SD of the distribution. Our investigation in Section C of the Supplement revealed that changes in both zBMI parameters could affect both the resulting sample statistics, suggesting the method may benefit from a more sophisticated multivariate optimization scheme. However, since we managed to achieve a high level of convergence using our na{\"i}ve scheme (100\% for percentile and 90\% for BMI), we anticipate that any improvements would be limited to the efficiency and reliability of the method rather than accuracy.

We have discussed a number of ways our methods can be improved and refined. In future work, we plan to develop this work further and explore the utility of our methods in other applications such as mapping between different depression scales. However, we recognise that even the best mapping method is not a substitute for data reported directly in trials. Our motivating example highlights that, in order to synthesize evidence about childhood weight status, trials are most useful when they report arm-level data on all BMI scales (un-standardized, zBMI and percentile). Trial reports should state the precise methods used to perform standardization, along with the relevant reference chart. Since childhood BMI is correlated with age and sex, it would be desirable for trials to describe the distribution of ages and the proportion of males and females in each intervention arm. It would also be useful for trials to report on the observed distribution of each outcome scale in their population. This would improve our understanding of the different scales and help to map between outcomes in historical trials.

\section{Summary}

To the best of our knowledge, this study is the first attempt to develop methods to map between arm-level aggregate data on different measurement scales. Previous techniques have involved simple re-scaling of contrast-level data via standardization and ratio techniques, or by fitting approximate models. In contrast, our methods leverage known transformations between scales at the individual level. The sampling and optimization methods possess broad applicability for mapping between any pair of measurement scales with known or estimable relationships between individual measurements. These mapping techniques enhance the potential for meta-analysis by facilitating the creation of larger datasets, enabling a more comprehensive summary of evidence from trials and improving the precision of estimates.

%\backmatter
\section*{Author contributions}
Annabel L Davies: writing - original draft (lead), formal analysis (lead), methodology (equal), software (lead), visualization (lead), writing - review \& editing (equal). A E Ades: conceptualization (equal), methodology (equal), writing - review \& editing (equal). Julian PT Higgins: conceptualization (equal), funding acquisition (lead), writing - review \& editing (equal).

\section*{Acknowledgments}
ALD and JPTH acknowledge funding from the National Institute for Health and Care Research (NIHR) Public Health Research programme (NIHR131572). JPTH is an NIHR Senior Investigator.

\section*{Data availability statement}
The data and code that support the findings of this paper are provided in the GitHub repository here: \url{https://github.com/AnnieDavies/Mapping_BMI}

\section*{Financial disclosure}

None reported.

\section*{Conflict of interest}

The authors declare no potential conflict of interests.

\section*{Supporting information}
Supplementary material is provided online.

\section*{Highlights}
\noindent\textbf{What is already known:}
\begin{itemize}
    \item Meta-analysis of continuous data is difficult when trials report aggregate outcomes on different measurement scales.
    \item Existing methods to synthesize evidence on different scales rely on simplifying assumptions such as linearity and non-negative outcomes.
\end{itemize}
\noindent\textbf{What is new:}
\begin{itemize}
    \item We develop three methods (analytical, sampling and optimization) for mapping between arm-level aggregate data on different measurement scales.
    \item We validate and compare the methods using trials that report childhood obesity outcomes on different body mass index (BMI) scales.
\end{itemize}
\noindent\textbf{Potential impact for readers:}
\begin{itemize}
    \item The sampling and optimization methods proposed can be used to map between aggregate data on any pair of measurement scales, provided we know, or can estimate, the relationship between individual measurements on the different scales. 
    \item Therefore, the methods facilitate the creation of larger, more comprehensive datasets for meta-analysis.
\end{itemize}

\bibliography{bibliography}

\appendix

\section{Smooth updating scheme for the optimization method}\label{app:update}
In this section we derive the updating scheme implemented in the optimization method (Equation (\ref{eq:update})). Let us assume that, at iteration $t$, zBMI is sampled from a normal distribution with mean $m_z^{(t)}$ and standard deviation $s_z^{(t)}$. Then, at iteration $t+1$, the mean and SD are increased by $\delta_m$ and $\delta_s$ respectively. The resulting sampling distributions at each iteration are
\begin{align}
	Z_{i}^{(t)} &\sim \mathcal{N}(m_z^{(t)}, s_z^{(t)2}),\\
	Z_{i}^{(t+1)} &\sim \mathcal{N}(m_z^{(t)}+\delta_m, (s_z^{(t)}+\delta_s)^2).
\end{align}
To write $Z_{i}^{(t+1)}$ as a function of $Z_{i}^{(t)}$ we transform both variables to a standard normal and equate. If a random variable $X$ is normally distributed via $X\sim \mathcal{N}(m,s^2)$, then the transformed variable $\frac{X-m}{s}$ follows a standard normal. Therefore,
\begin{align}
	\frac{Z_{i}^{(t)}-m_z^{(t)}}{s_z^{(t)}} &\sim \mathcal{N}(0, 1), \label{eq:zt}\\
	\frac{Z_{i}^{(t+1)}-(m_z^{(t)}+\delta_m)}{s_z^{(t)}+\delta_s} &\sim \mathcal{N}(0, 1). \label{eq:zt1}
\end{align}
Equating the left hand side of Equations (\ref{eq:zt}) and (\ref{eq:zt1}) leads to
\begin{align}
	\frac{Z_{i}^{(t)}-m_z^{(t)}}{s_z^{(t)}} = \frac{Z_{i}^{(t+1)}-(m_z^{(t)}+\delta_m)}{s_z^{(t)}+\delta_s}
\end{align}
which, rearranging for $Z_{i}^{(t+1)}$, gives the result quoted in Section \ref{sec:opt} of the main paper,
\begin{align}
	Z_i^{(t+1)} = \left( 1 + \frac{\delta_{s}}{s_z^{(t)}} \right) (Z_i^{(t)}- m_z^{(t)}) + (m_z^{(t)} + \delta_{m}).
\end{align}

%\nocite{*}% Show all bib entries - both cited and uncited; comment this line to view only cited bib entries;

\end{document}

\renewcommand{\thepage}{S\arabic{page}}
 \setcounter{page}{1}

\author{Annabel L. Davies}
\author{A E Ades}
\author{Julian PT Higgins}

\authormark{DAVIES et al}

\noindent{\titlefont Mapping between measurement scales in meta-analysis, with application to measures of body mass index in children}

\noindent\ifnum\aucount>0%
   \global\punctcount\aucount%
   {\artauthors\par}%
   \removelastskip\vskip8.25pt%
%   {\jmkaddress\par}%
\fi%

\appendix

\section{Reference charts for WHO and IOTF}\label{app:charts}
\begin{figure*}[h]
	\centering
	\includegraphics[width=1\linewidth]{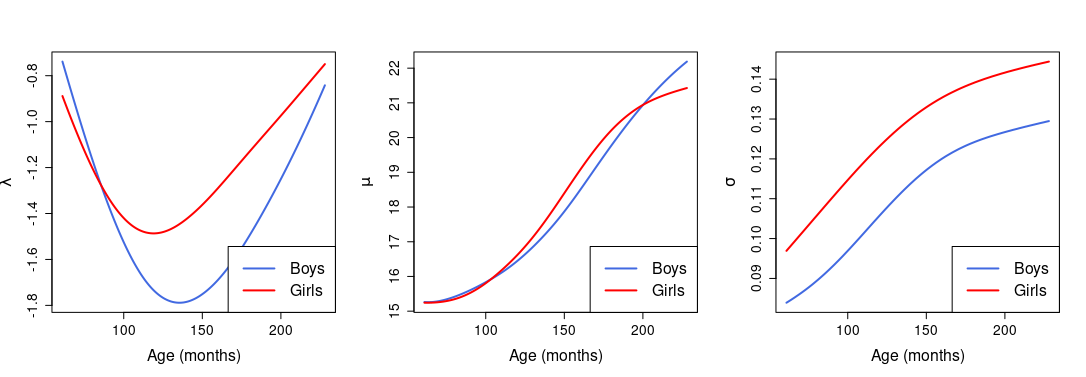}
	\caption{WHO (World Health Organization) reference parameters $\{\lambda, \mu, \sigma\}$ as a function of age. Parameter values were obtained from Reference\cite{WHO} (Online).}
	\label{fig:WHO-LMS}
\end{figure*}

\begin{figure*}[h]
	\centering
	\includegraphics[width=1\linewidth]{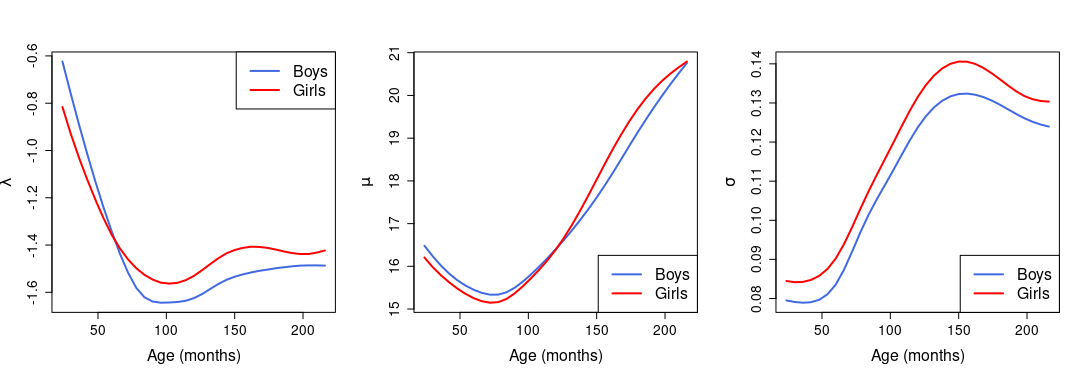}
	\caption{IOTF (International Obesity Taskforce) reference parameters $\{\lambda, \mu, \sigma\}$ as a function of age. Parameter values obtained from Reference\cite{IOTF} (Supplementary Table 1).}
	\label{fig:IOTF-LMS}
\end{figure*}

Figures \ref{fig:WHO-LMS} and \ref{fig:IOTF-LMS} show the LMS parameters for boys and girls as a function of age provided by the World Health Organization (WHO)\cite{WHO} and the International Obesity Taskforce (IOTF),\cite{IOTF} respectively. Both charts are based on surveys across six countries aiming to represent the international population. The WHO chart is based on data from Brazil, Ghana, India, Norway, Oman and the US\cite{WHO:survey:2006,deOnis:2006} whereas the IOTF survey covered Brazil, Great Britain, Hong Kong, the Netherlands, Singapore, and the US.\cite{Cole:2000}
\vspace{-20pt}
\vspace{-20pt}

\section{Assuming a beta distribution for percentile BMI}\label{app:beta}
To perform the mapping $\mathcal{D}_{P}  \xrightarrow{} \mathcal{D}_Z$ using the sampling method we assume that percentile follows a beta distribution,
\begin{align}
	P \sim \text{Beta}(\alpha, \beta).
\end{align}
In this section we show that this assumption yields a normal or approximately normal distribution for zBMI under certain parameterisations. 

\begin{figure*}[b]
	\centering
	\includegraphics[width=0.55\linewidth]{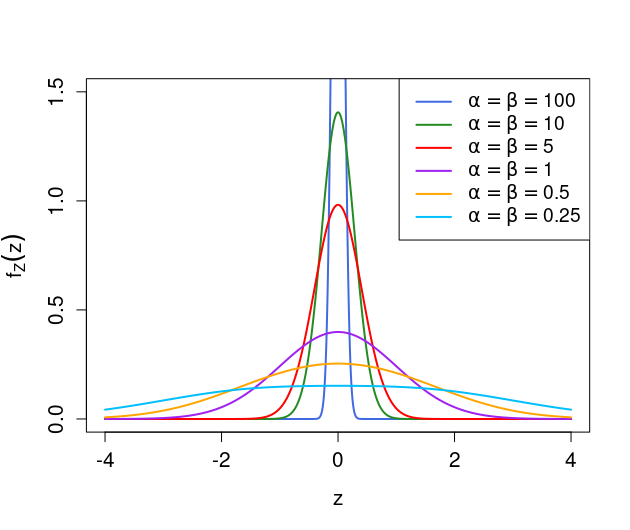}
	\caption{The probability density function of zBMI, $f_Z(.)$, evaluated using Equation (\ref{eq:Zdist}) and assuming the pdf of percentile, $f_P(.)$, describes a beta distribution, $\text{Beta}(\alpha, \beta)$. We show the result for various parameterisations of the percentile distribution where $\alpha=\beta$. }
	\label{fig:Zdist}
\end{figure*}

First, we derive an expression for the distribution of $Z$ in terms of the distribution of $P$. We use the fact that $Z = \Phi^{-1}(P)$ to write the cumulative distribution function (cdf) of $Z$ as
\begin{align}
	F_{Z}(z) &= \text{Pr}(Z \leq z) \nonumber\\
	&= \text{Pr}(\Phi^{-1}(P)\leq z)\nonumber\\
	&= \text{Pr}(P \leq \Phi(z)) \nonumber\\
	&= F_P(\Phi(z)),
\end{align} 
where $F_P(.)$ is the cdf of percentile. We now obtain the probability density function (pdf) of $Z$ by taking the derivative of the cdf,
\begin{align}
	f_Z(z) &= \frac{\text{d}}{\text{d}z} F_Z(z)\nonumber \\
	&= \frac{\text{d}}{\text{d}z} F_P(\Phi(z))\nonumber\\
	&=\frac{\text{d}}{\text{d}\Phi(z)} F_P(\Phi(z))  \frac{\text{d}}{\text{d}z} \Phi(z)\nonumber\\
	&=f_P(\Phi(z))\phi(z), \label{eq:Zdist}
\end{align}
where $f_P(.)$ is the pdf of percentile, which we have defined as a beta distribution, and $\phi(.)$ is the pdf of the standard normal. As shown in Figure \ref{fig:Zdist}, setting $\alpha=\beta$ in the beta distribution, $f_P(.)$, yields a pdf of $Z$ that resembles a normal distribution.

\begin{figure*}[h]
	\centering
	\includegraphics[width=1\linewidth]{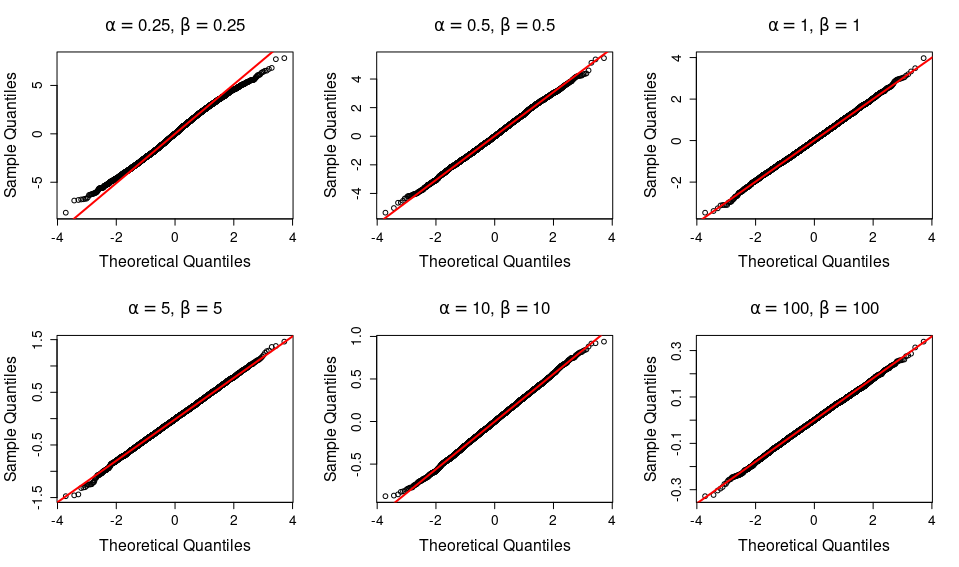}
	\caption{Q-Q plots comparing the distribution of $Z$ with a normal distribution where $Z$ is generated from $Z=\Phi^{-1}(P)$ assuming $P\sim \text{Beta}(\alpha, \beta)$ for different values of $\alpha=\beta$. }
	\label{fig:QQ}
\end{figure*}

To verify the distributions in Figure \ref{fig:Zdist} are normal (or sufficiently close to normal), we use a Q-Q (quantile-quantile) plot.\cite{Wilk:1968} We sample values of $P$ from a beta distribution and convert them to zBMI using $Z=\Phi^{-1}(P)$. We then feed the corresponding values of $Z$ into the R function \texttt{qqnorm()}. If the resulting pdf $f_Z(.)$ is a normal distribution, then the data points from \texttt{qqnorm()} will follow a straight line. In Figure \ref{fig:QQ} we find that for $\alpha=\beta \geq 0.5$, the plotted quantiles closely follow the straight line. Therefore, under certain parameterizations, assuming a beta distribution for percentile yields a normal distribution for zBMI.

\section{Selecting optimization parameters for convergence}\label{app:Converge}
To implement the optimization method and ensure convergence we need to choose reasonable values for the incremental step, $\Delta_{\text{step}}$, and tolerance threshold, $\Delta_{\text{tol}}$. In the following sections we investigate the convergence of algorithms 3 and 4 for various values of these parameters. To choose a reasonable space of values for investigation, we consider typical values of the different scales and explore how changes in the parameters of the zBMI distribution lead to changes in the other scales.

\subsection{Algorithm 3: zBMI from percentile}

In the general population, percentile follows a uniform distribution between 0 and 1. Therefore, a typical value for $\bar{P}$ is 0.5, while SDs are typically about 0.3. We choose to investigate $\Delta_{\text{tol}}=(0.0005, 0.001, 0.005)$ which correspond to typical percentage errors of (0.1\%, 0.2\%, 1\%) on $\bar{P}$
and (0.2\%, 0.3\%, 2\%) on $\text{SD}(P)$.

The incremental step, $\Delta_{\text{step}}$, characterizes changes in the parameters of the zBMI distribution. Since $Z$ ranges from approximately -3 to 3 while $P$ is restricted to the range 0 to 1, changes in zBMI parameters lead to smaller changes on the percentile scale. In numerical tests, we found that increasing (decreasing) $m_z$ increased (decreased) the mean of percentile. The ratio of change in $P$ to change in $Z$ varied from 0.05 to 0.25 depending on the value of $m_z$. Similarly, increasing (decreasing) $s_z$ increased (decreased) the SD of percentile. Here, the ratio of changes varied from 0.05 to 0.4 depending on the value of $s_z$. Changes in $s_z$ had no effect on the mean of percentile but changes in $m_z$ did cause changes in $\text{SD}(P)$. For $m_z<0$, increasing (decreasing) $m_z$ increased (decreased) $\text{SD}(P)$ (ratio of changes varied from 0 to 0.1). For $m_z>0$, changes occurred in the opposite direction such that increasing (decreasing) $m_z$ decreased (increased) $\text{SD}(P)$ (ratio of changes varied from 0 to 0.1). 

If the incremental step and threshold were on the same scale, we would choose $\Delta_{\text{step}} \approx 0.5 \Delta_{\text{tol}}$. Because these values are not on the same scale the choice is more difficult. Based on our numerical tests we chose to investigate values of $\Delta_{\text{step}}$ which give the following ratios, $\Delta_{\text{step}}/\Delta_{\text{tol}} = (1/5, 2/5, 3/5)$. Table \ref{Tab:Alg3} shows the full space of $(\Delta_{\text{step}},\Delta_{\text{tol}})$ values investigated, and the percentage of estimates that converged within $n_{\text{max}}=5000$ iterations. For all sets of  $(\Delta_{\text{step}},\Delta_{\text{tol}})$ with over 85\% convergence, we observed negligible differences in the estimates of $\mathcal{D}_Z$. In the main paper we show the results for $\Delta_{\text{step}}=0.002$ and $\Delta_{\text{tol}}=0.005$ (100\% convergence).

\begin{center}
\begin{table*}[h]
\caption{Convergence status of Algorithm 3 for different optimization parameters.\label{Tab:Alg3}}
\begin{tabular*}{\textwidth}{@{\extracolsep\fill}lllllll@{}}
\toprule
&\multicolumn{6}{@{}l}{$\Delta_{\text{step}}/\Delta_{\text{tol}}$} \\\cmidrule{2-7}
&\multicolumn{2}{@{}l}{\textbf{1/5}} & \multicolumn{2}{@{}l}{\textbf{2/5}} &  \multicolumn{2}{@{}l}{\textbf{3/5}} \\\cmidrule{2-3}\cmidrule{4-5}\cmidrule{6-7}
$\Delta_{\text{tol}}$ & $\Delta_{\text{step}}$ & \% converged & $\Delta_{\text{step}}$ & \% converged & $\Delta_{\text{step}}$ & \% converged  \\
\midrule
0.0005 & 0.0001 & 17\% & 0.0002 & 67\% & 0.0003 & 87\% \\ 
0.001 & 0.0002 & 62\% & 0.0004 & 98\% & 0.0006 & 100\% \\ 
0.005 & 0.001 & 100\% & 0.002 & 100\% & 0.003 & 100\% \\ 
\bottomrule
\end{tabular*}
\end{table*}
\end{center}

\subsection{Algorithm 4: zBMI from BMI}
In our data set, a typical value of $\bar{B}$ is 20 $\text{kg}/\text{m}^2$, with SDs of around 5 $\text{kg}/\text{m}^2$. Therefore, we chose to investigate $\Delta_{\text{tol}}=(0.01, 0.05, 0.1)$ which correspond to typical percentage errors of (0.05\%, 0.25\%, 0.5\%) on $\bar{B}$ and (0.2\%, 1\%, 2\%) on $\text{SD}(B)$.

As before, the incremental step $\Delta_{\text{step}}$ corresponds to changes in the parameters of the zBMI distribution. In our data set, mean BMI ranges from approximately 15 to 28 $\text{kg}/\text{m}^2$. Therefore, changes in zBMI parameters lead to larger changes on the BMI scale. In numerical tests, we found that increasing (decreasing) $m_z$ increased (decreased) the mean of BMI. The ratio of change in $B$ to change in $Z$ varied from approximately 2 to 5 depending on the value of $m_z$. Similarly, increasing (decreasing) $s_z$ increased (decreased) the SD of BMI. Here, the ratio of change varied from approximately 1 to 6 depending on the value of $s_z$. Changes in $m_z$ had minimal effect on the SD of BMI but changes in $s_z$ did cause (small) changes in $\bar{P}$. On average, we observed that increasing (decreasing) $s_z$ increased (decreased) $\bar{P}$ (ratio varied from 0 to 3). 

Based these tests we chose investigate values of $\Delta_{\text{step}}$ which give the following ratios, $\Delta_{\text{step}}/\Delta_{\text{tol}} = (1/10, 1/5, 1/2)$. Table \ref{Tab:Alg4} shows the percentage of estimates that reached convergence within $n_{\text{max}}=5000$ iterations. The results for all sets of $(\Delta_{\text{step}},\Delta_{\text{tol}})$ showed negligible differences in the estimates of $\mathcal{D}_Z$ suggesting estimates that did not converge were still close to their converged values. In the main paper we show the results for $\Delta_{\text{step}}=0.01$ and $\Delta_{\text{tol}}=0.1$ (highest convergence rate, 90\%).

\begin{center}
\begin{table*}[!h]%
\caption{Convergence status of Algorithm 4 for different optimization parameters.\label{Tab:Alg4}}
\begin{tabular*}{\textwidth}{@{\extracolsep\fill}lllllll@{}}
\toprule
&\multicolumn{6}{@{}l}{$\Delta_{\text{step}}/\Delta_{\text{tol}}$} \\\cmidrule{2-7}
&\multicolumn{2}{@{}l}{\textbf{1/5}} & \multicolumn{2}{@{}l}{\textbf{2/5}} &  \multicolumn{2}{@{}l}{\textbf{3/5}} \\\cmidrule{2-3}\cmidrule{4-5}\cmidrule{6-7}
$\Delta_{\text{tol}}$ & $\Delta_{\text{step}}$ & \% converged & $\Delta_{\text{step}}$ & \% converged & $\Delta_{\text{step}}$ & \% converged  \\
\midrule
0.01 & 0.001 & 89\% & 0.002 & 83\% & 0.005 & 56\% \\ 
0.05 & 0.005 & 88\% & 0.01 & 83\% & 0.025 & 60\% \\ 
0.1 & 0.01 & 90\% & 0.02 & 85\% & 0.05 & 58\% \\ 
\bottomrule
\end{tabular*}
\end{table*}
\end{center}

\section{Comparing the effect of age distributions on mapping from BMI}\label{app:Age}
\begin{figure*}[h!]
	\centering
	\includegraphics[width=1\linewidth]{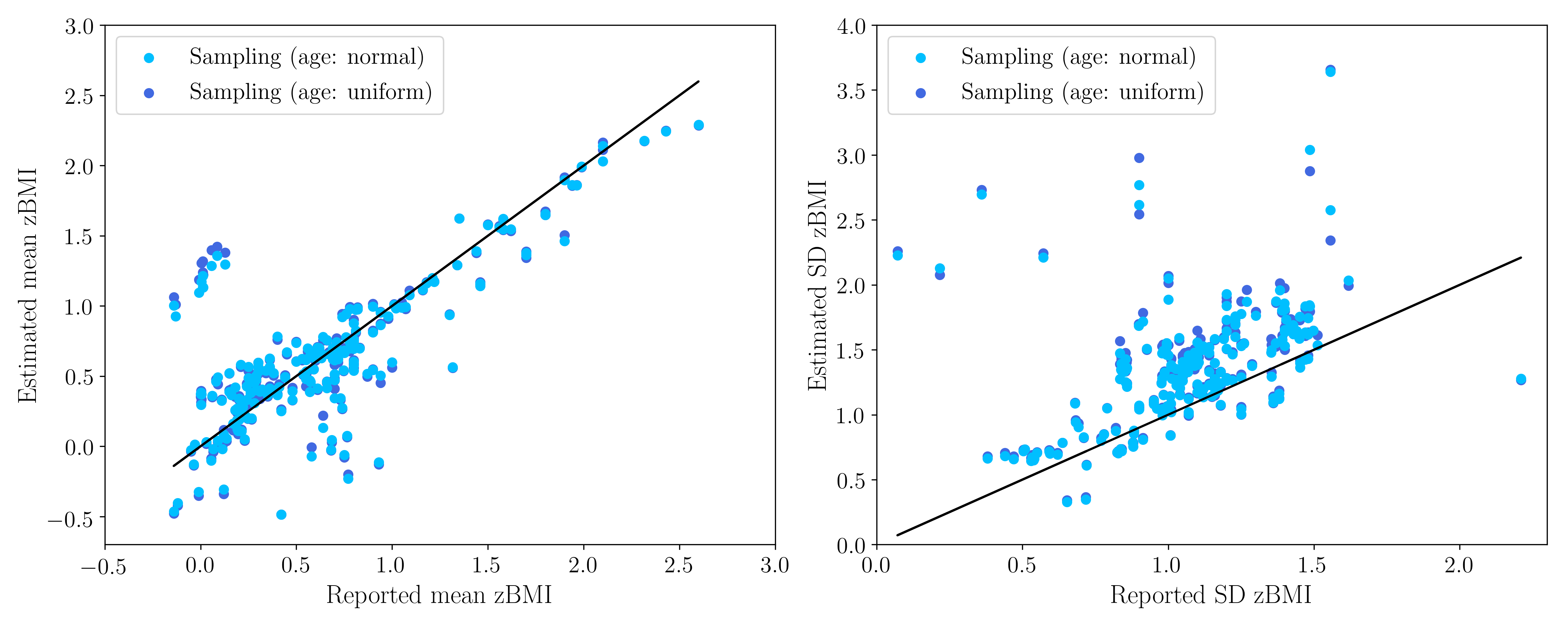}
	\caption{A plot comparing the effect of different age distributions (normal and uniform) on the sampling method for mapping from BMI to zBMI (algorithm 2). }
	\label{fig:CompareAge}
\end{figure*}

Figure \ref{fig:CompareAge} shows the results for Algorithm 2 (sampling method for mapping from BMI to zBMI) when assuming a uniform distribution for age compared with a normal distribution. Table \ref{Tab:Age} shows the corresponding values of RMSE and MAE. The two distributions exhibit similar errors, but the normal distribution slightly outperforms the uniform distribution.

\begin{center}
\begin{table*}[!h]%
\caption{The root mean squared error (RMSE) and mean absolute error (MAE) on the estimated aggregate data (mean and SD) from the BMI sampling method assuming different distributions for age.\label{Tab:Age}}
\begin{tabular*}{\textwidth}{@{\extracolsep\fill}lllll@{}}
\toprule
&\multicolumn{2}{@{}l}{\textbf{Estimated mean zBMI}} & \multicolumn{2}{@{}l}{\textbf{Estimated SD of zBMI}} \\\cmidrule{2-3}\cmidrule{4-5}
\textbf{Method} & \textbf{RMSE}  & \textbf{MAE}  & {\textbf{RMSE}}  & \textbf{MAE}   \\
\midrule
Sampling (age: normal) & 0.354 & 0.218 &  0.504 &  0.336  \\ 
Sampling (age: uniform) & 0.368 & 0.221 & 0.516 & 0.349 \\
\bottomrule
\end{tabular*}
\end{table*}
\end{center}

\section{Exploring outliers}\label{app:outliers}
\begin{figure*}[h!]
	\centering
	\includegraphics[width=1\linewidth]{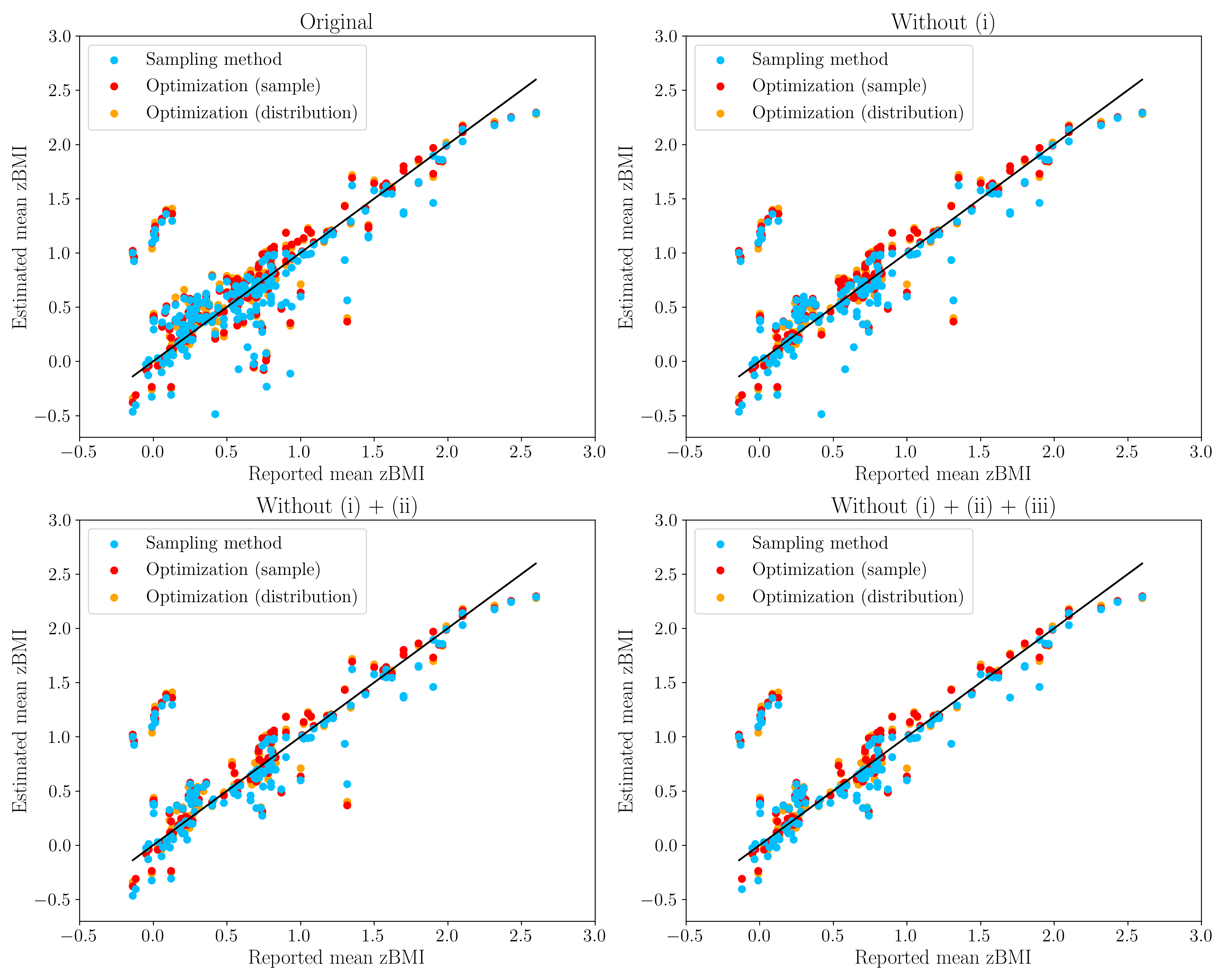}
	\caption{Plots showing how the mean zBMI results for $\mathcal{D}_B \xrightarrow{} \mathcal{D}_Z$ are impacted by the following assumptions or approximations:(i) no reported LMS reference chart, (ii) SDs which we unadjusted for clustering, and (iii) follow-up values imputed from change scores. The top left plot shows the results for all data (this is the same as the left panel of Figure 3 in the main text but without the distribution estimates from the optimization method). The top right panel shows the same results without data that are subject to condition (i). The bottom left panel shows the results without data corresponding to either condition (i) or (ii). Finally, the bottom right panel shows the plot removing data subject any of the three conditions. }
	\label{fig:Outliers-mean}
\end{figure*}

\begin{figure*}[h!]
	\centering
	\includegraphics[width=1\linewidth]{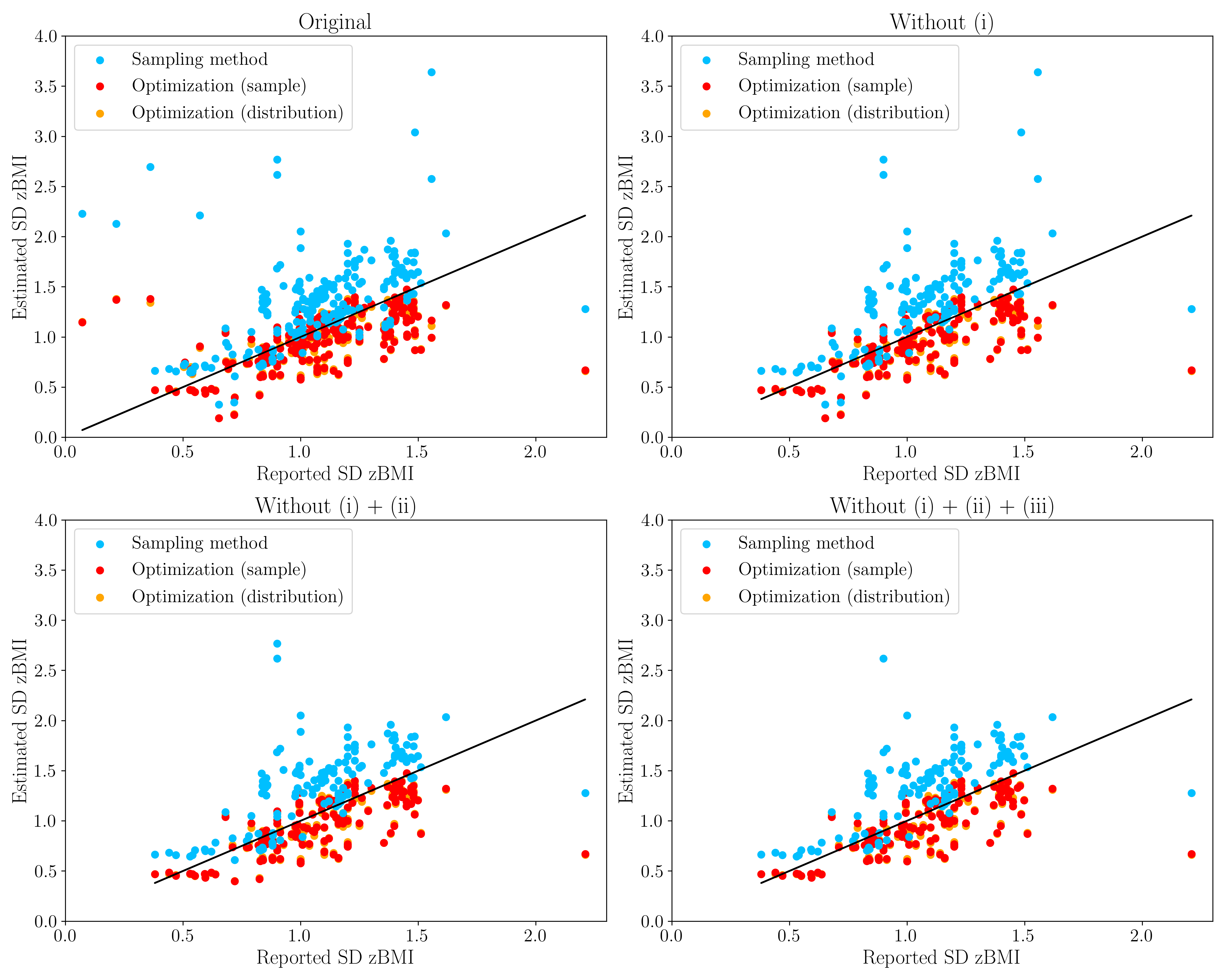}
	\caption{Plots showing how the SD zBMI results for $\mathcal{D}_B \xrightarrow{} \mathcal{D}_Z$ are impacted by the following assumptions or approximations:(i) no reported LMS reference chart, (ii) SDs which we unadjusted for clustering, and (iii) follow-up values imputed from change scores. The top left plot shows the results for all data (this is the same as the right panel of Figure 3 in the main text but without the distribution estimates from the optimization method). The top right panel shows the same results without data that are subject to condition (i). The bottom left panel shows the results without data corresponding to either condition (i) or (ii). Finally, the bottom right panel shows the plot removing data subject any of the three conditions. }
	\label{fig:Outliers-SD}
\end{figure*}

To try and understand some of the outliers in the mapping results for $\mathcal{D}_B \xrightarrow{} \mathcal{D}_Z$, we removed data from the plot which were subject to the following assumptions or approximations: (i) no reported LMS reference chart, (ii) SDs that we unadjusted for clustering, and (iii) follow-up values imputed from change scores. Figure \ref{fig:Outliers-mean} shows the mean zBMI results for the sampling and optimization methods excluding data corresponding to (i), (i) + (ii), and (i)+(ii)+(iii). Figure \ref{fig:Outliers-SD} shows the same for standard deviation. Removing all three conditions reduces the scatter on both the mean and SD plots. Except for the cluster of outliers corresponding to Papadaki (2010), the mapped mean zBMI values now exhibit good agreement with the reported values. The right-most data point on each plot in Figure \ref{fig:Outliers-SD} corresponds to Brown (2013) which, as discussed in Section 5.2.1 of the main paper, we suspect has been misreported in the trial. Therefore, except for a couple of outliers, the standard deviation results also show reasonable agreement when accounting for conditions (i) to (iii).

\clearpage
\section{Exploring reference charts}\label{app:sep-charts}

\begin{figure*}[h!]
	\centering
	\includegraphics[width=1\linewidth]{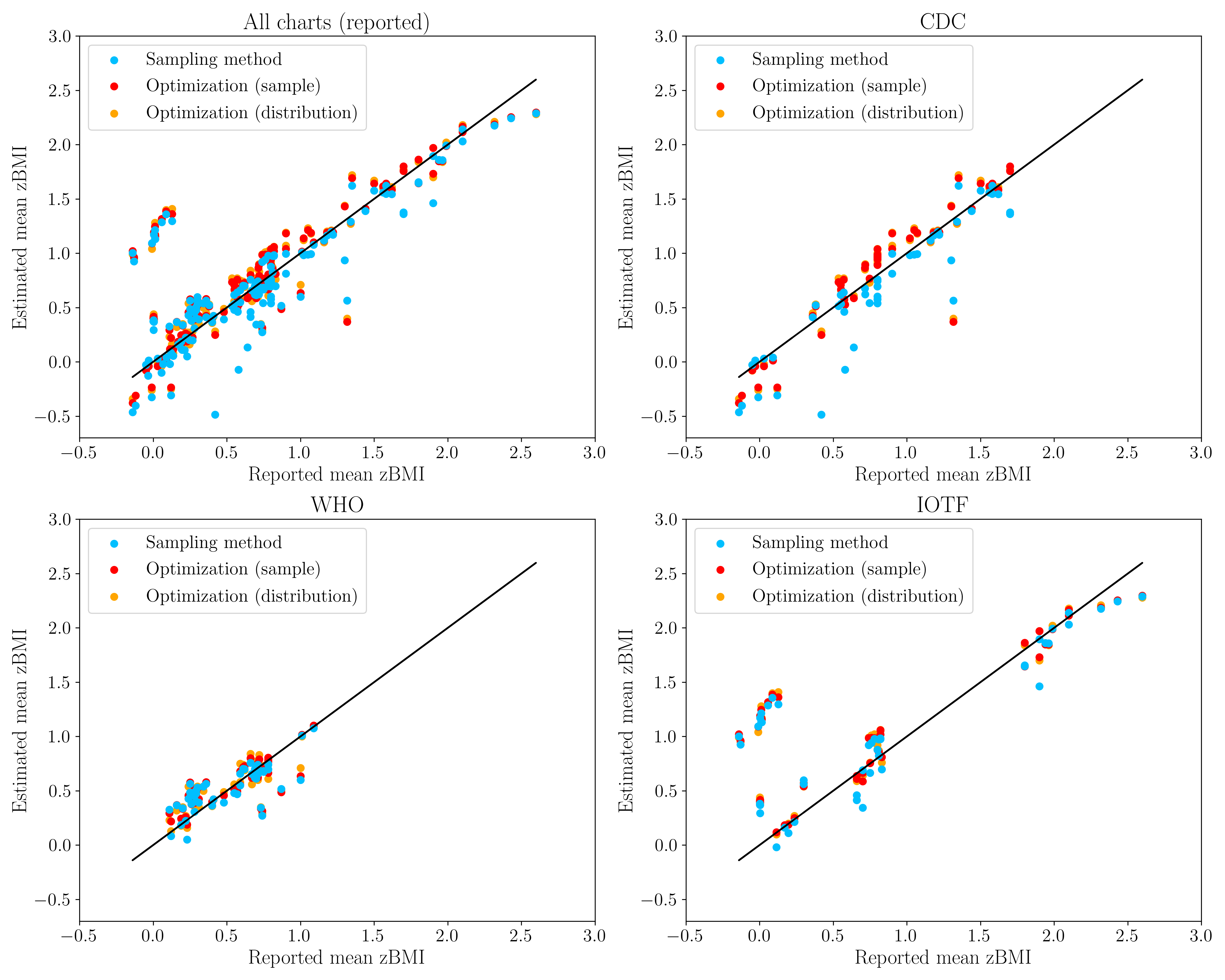}
	\caption{Plots showing the mean zBMI results for $\mathcal{D}_B \xrightarrow{} \mathcal{D}_Z$ stratified by the LMS reference chart used in the mapping. The top left panel shows the results excluding data where the reference chart is not reported (this is the same as the top right panel in Figure \ref{fig:Outliers-mean}). The top right panel shows only the results for trials which report using the CDC reference chart. The bottom left panel shows the results for the WHO reference chart and the bottom right panel shows the plot for the IOTF chart. }
	\label{fig:Charts-mean}
\end{figure*}

\begin{figure*}[h!]
	\centering
	\includegraphics[width=1\linewidth]{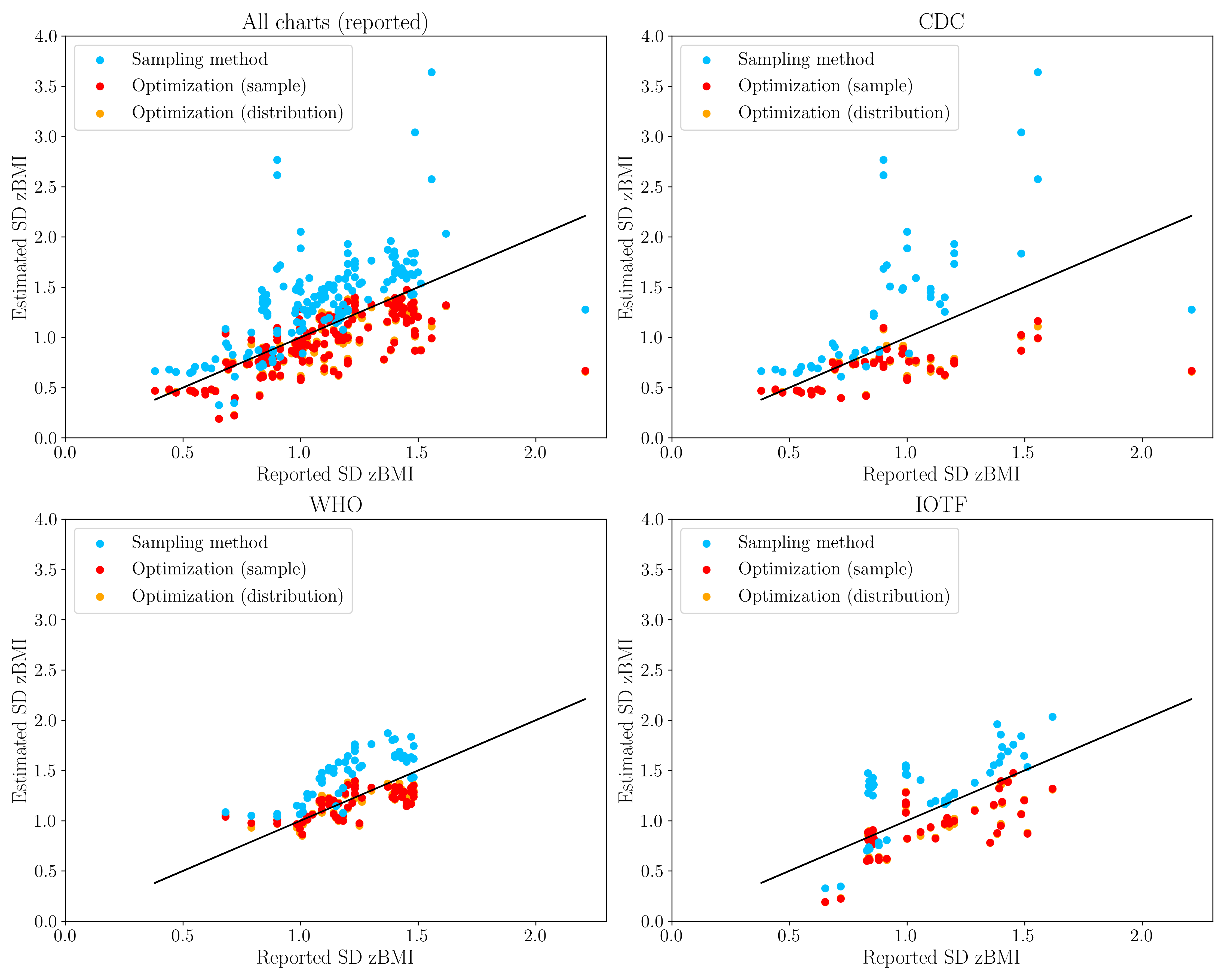}
	\caption{Plots showing the SD zBMI results for $\mathcal{D}_B \xrightarrow{} \mathcal{D}_Z$ stratified by the LMS reference chart used in the mapping. The top left panel shows the results excluding data where the reference chart is not reported (this is the same as the top right panel in Figure \ref{fig:Outliers-SD}). The top right panel shows only the results for trials which report using the CDC reference chart. The bottom left panel shows the results for the WHO reference chart and the bottom right panel shows the plot for the IOTF chart.}
	\label{fig:Charts-SD}
\end{figure*}

For the BMI optimization method, we investigate the effect of the reference chart used to obtain the LMS parameters. In Figures \ref{fig:Charts-mean} and \ref{fig:Charts-SD} we show the estimated mean and SD for trials which report each of the three main reference charts, CDC, WHO and IOTF. We exclude results from trials that do not report a reference chart. We do not observe any notable differences in the accuracy of the mean zBMI estimates between the charts. However, the estimates of SD are more scattered for the CDC chart compared with the others.

\clearpage
\section{Converged estimates only}\label{app:converge-only}
In Figure 3 in the main paper we show the results of the BMI optimization method (Algorithm 4) excluding the 10\% of estimates that did not converge. Except for the removal of the outlier on the far right of the SD plot (right panel), the results are very similar to the full dataset in Figure 3. Therefore, convergence issues do not seem to be the cause of the observed inaccuracies. 

\begin{figure*}[h!]
	\centering
	\includegraphics[width=1\linewidth]{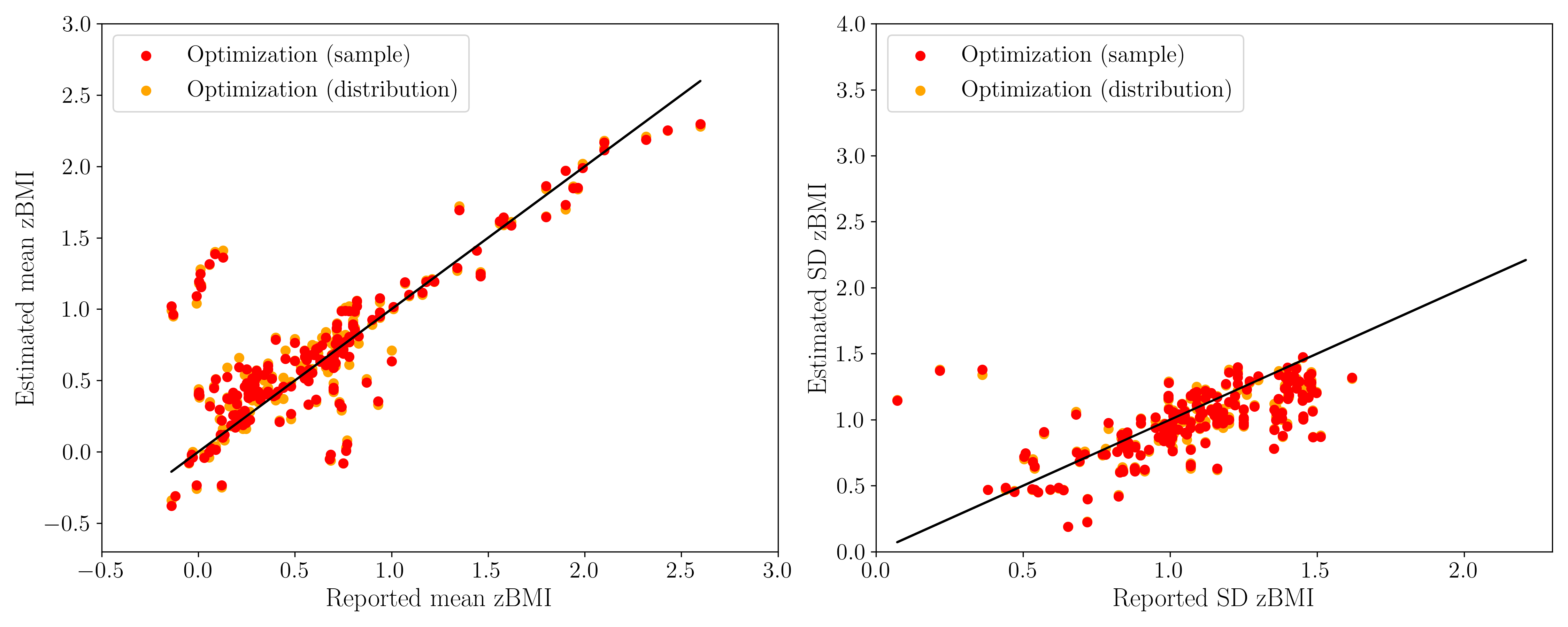}
	\caption{A plot showing the results of the optimization mapping methods for estimating aggregate zBMI data from aggregate BMI data excluding results that did not converge.}
	\label{fig:ZfromB-converged}
\end{figure*}

\bibliography{bibliography}%